\documentclass[authorversion, nonacm, manuscript, screen]{acmart}

\usepackage[english]{babel}
\usepackage{amsmath}
\usepackage{graphicx}
\usepackage{textcomp}
\usepackage{xcolor}
\usepackage{tikz} 
\usepackage{pgfplots} 
\usepackage{float}
\usepackage{longtable}
\usepackage{multirow}
\usepackage{algorithm}
\usepackage{algpseudocode}
\usepackage{booktabs}
\usepackage{todonotes}
\usepackage{soul}		
\usepackage{dblfloatfix}
\usepackage{subcaption}
\usepackage{nameref}
\usepackage{todonotes}
\usepackage{hyperref}
\usepackage{tabularx}
\usepackage{multirow}
\usepackage{tablefootnote}
\usepackage{adjustbox}

\def\BibTeX{{\rm B\kern-.05em{\sc i\kern-.025em b}\kern-.08em
		T\kern-.1667em\lower.7ex\hbox{E}\kern-.125emX}}

\graphicspath{{./img/}}

\usepgfplotslibrary{units}
\usetikzlibrary{pgfplots.groupplots}

\pgfplotsset{compat = 1.3}

\newcommand{\ifm}{\textit{ifm} }
\newcommand{\ofm}{\textit{ofm} }

\definecolor{my_green}{RGB}{0,158,115}
\definecolor{my_blue}{RGB}{86,180,233}
\definecolor{my_orange}{RGB}{230,159,0}
\definecolor{my_yellow}{RGB}{240,228,66}
\definecolor{my_red}{RGB}{213,94,0}
\definecolor{my_purple}{RGB}{204,121,167}

\ccsdesc[500]{Computer systems organization~Distributed architectures}
\ccsdesc[500]{Computing methodologies~Distributed artificial intelligence}
\ccsdesc[300]{Computer systems organization~Embedded systems}
\ccsdesc[300]{General and reference~Surveys and overviews}

\begin{document}
	
	\title{Embedded Distributed Inference of Deep Neural Networks: A Systematic Review}
	
	\author{Federico Nicolás Peccia}
	\orcid{0000-0002-3587-0415}
	\affiliation{%
		\institution{FZI Research Center for Information Technology}
		\city{Karlsruhe}
		\country{Germany}
	}
	\email{peccia@fzi.de}
	
	\author{Oliver Bringmann}
	\orcid{0000-0002-1615-507X}
	\affiliation{%
		\institution{University of Tübingen}
		\city{Tübingen}
		\country{Germany}
	}
	\email{oliver.bringman@uni-tuebingen.de}
	
	\begin{abstract}
		Embedded distributed inference of Neural Networks has emerged as a promising approach for deploying machine-learning models on resource-constrained devices in an efficient and scalable manner. The inference task is distributed across a network of embedded devices, with each device contributing to the overall computation by performing a portion of the workload. In some cases, more powerful devices such as edge or cloud servers can be part of the system to be responsible of the most demanding layers of the network. As the demand for intelligent systems and the complexity of the deployed neural network models increases, this approach is becoming more relevant in a variety of applications such as robotics, autonomous vehicles, smart cities, Industry 4.0 and smart health. We present a systematic review of papers published during the last six years which describe techniques and methods to distribute Neural Networks across these kind of systems. We provide an overview of the current state-of-the-art by analysing more than 100 papers, present a new taxonomy to characterize them, and discuss trends and challenges in the field.
	\end{abstract}
	
	\keywords{Edge, distributed systems, Neural Networks}
	
	
	
	\maketitle
	
	
	
	\section{Introduction}
\label{section:introduction}

The execution of the inference pass of Deep Neural Networks (DNN) on systems composed of multiple devices presents advantages for a variety of use cases. For example, incorporating distributed inference in industrial automation can enable real-time monitoring of machines with physically distant sensors, enhancing efficiency by reducing communicated data and lowering downtime \cite{wangFastEnergySavingNeural2021}. Distributed inference can also improve privacy by keeping user-sensitive data close to the source that generates it and avoiding sharing raw data with a centralized server \cite{shiPrivacyAwareEdgeComputing2019,baccourRLPDNNReinforcementLearning2021a}. For smart city scenarios, distributed inference can be used to improve video analytics performance \cite{fuSplitComputingVideo2022a}. In smart health applications, this technique is used to improve the availability of system composed of multiple distributed healthcare monitors \cite{zhangAccelerateDeepLearning2020a}, or to aid geriatric care scenarios \cite{naveenLowLatencyDeep2021a}.

\begin{figure}[!t]
	\centering
	\includegraphics[width=\textwidth]{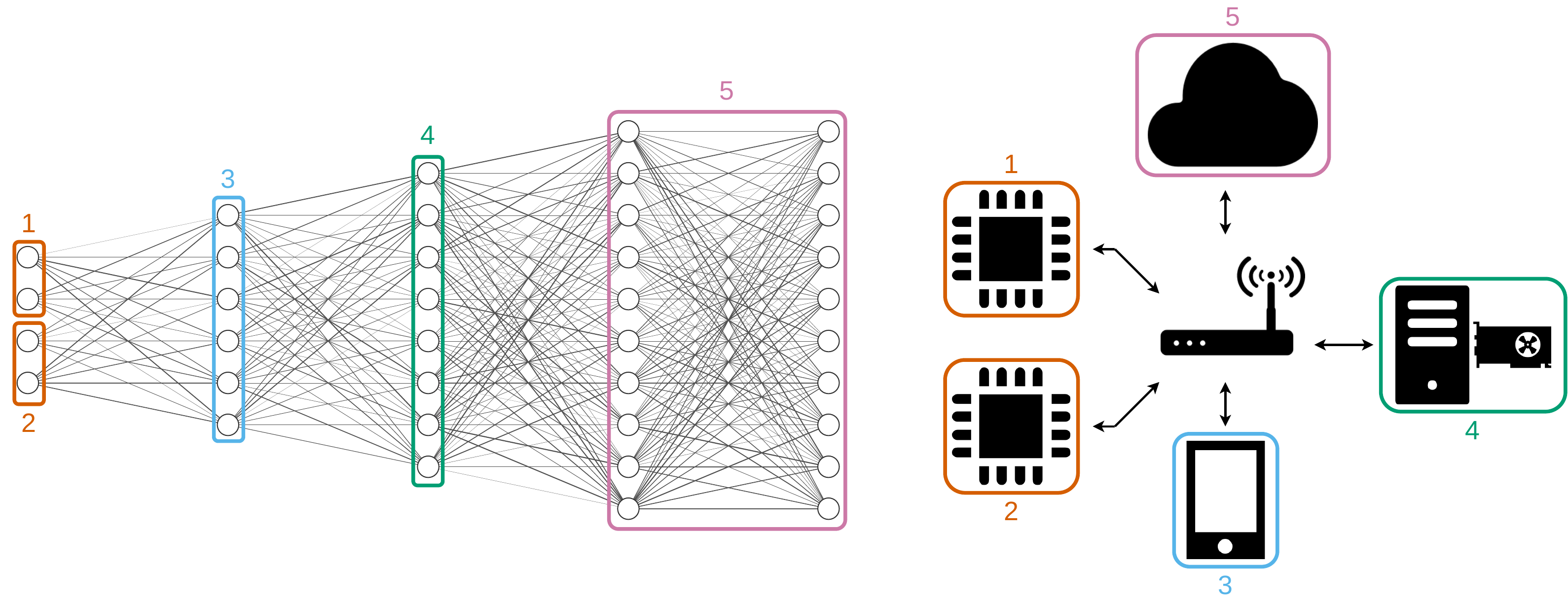}
	\Description{A schematic representation of the partitioning and allocation of a DNN across multiple devices. Left, a generic neural network can be seen. Each group of neurons is marked with a different colour rectangle. To the right, a network composed of embedded devices, edge and cloud servers shows with different colors which neurons are assigned for execution on each device.}
	\caption{A schematic representation of the partition and allocation of a DNN across multiple hardware components, including embedded devices parallelizing the execution of the same layer, edge ones running more complex layers, and cloud servers executing the more computation demanding layers.}
	\label{fig:distributed_inference_diagram}
\end{figure} 

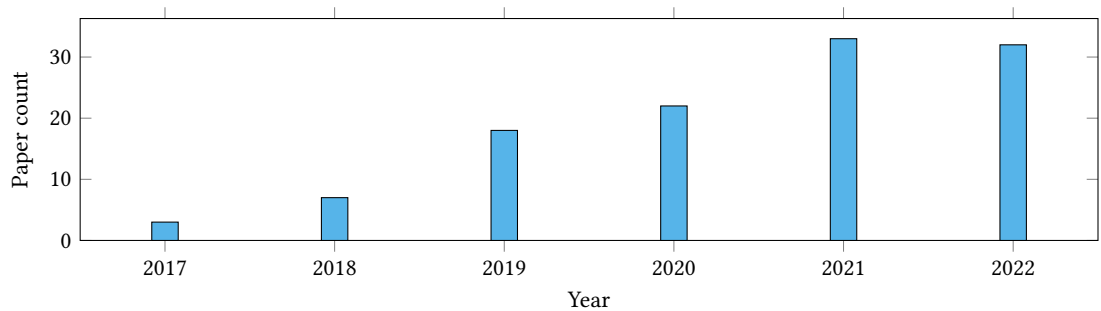
\begin{figure}[!ht]
	\centering
	\begin{tikzpicture}
		\begin{axis}[
			height=0.3\textwidth,
			width=\textwidth,
			xlabel=Year,
			ylabel=Paper count,
			ybar,
			ymin=0,
			xtick=data,
			symbolic x coords={
				2017,
				2018,
				2019,
				2020,
				2021,
				2022}]
			\addplot[fill=my_blue] table[x=Year,y=Count,col sep=comma] {./data/works_per_year.csv};
		\end{axis}
	\end{tikzpicture}
	\Description{A bar plot showing how many of the surveyed papers were published each year from 2018 to 2022. A clear ascending tendency can be appreciated, reflecting the growing interest of the research community on this topic.}
	\caption{Surveyed embedded distributed inference papers arranged per year}
	\label{fig:work_per_year}
\end{figure}

However, the continuously increasing memory footprint and computational complexity of current DNN architectures, together with hard constraints such as energy consumption and latency requirements, have motivated a growing interest in finding an efficient and automated distribution of the inference of an DNN across multiple devices (Figure\ref{fig:work_per_year}). Previous surveys, such as \cite{baccourPervasiveAIIoT2022,chenDistributedLearningWireless2021} have addressed the distribution of AI algorithms across multiple devices, but have focused on different aspects of the topic (federated learning, reinforcement learning, active learning, pervasive inference, privacy of distributed AI systems, etc.), thus dedicating less space to the particular problem of \textit{distributed inference}. In addition, neither of these surveys used the systematic review methodology; in their respective sections dedicated to distributed inference, \cite{baccourPervasiveAIIoT2022} and \cite{chenDistributedLearningWireless2021} only analyzed 36 and 41 papers, respectively. They also focused on low-level techniques of partitioning each type of layer, but dedicated little analysis to other aspects of these papers (problem definition, adaptability of the resulting distributed system, etc).

Contrary to previous surveys, this work focuses on multiple aspects of the techniques and methodologies used to achieve distributed inference, exploring how to partition a DNN and allocate the execution of each section across a variety of devices. We surveyed more than 100 papers that used distributed inference on embedded and edge devices and provided qualitative (categorizing them according to their characteristics: runtime flexibility, partition point granularity, optimization metrics, constraints, etc.) and quantitative analyses (comparing their reported metrics improvements between them). We also review the most commonly distributed DNN architectures, the typical embedded devices used in these studies and provide a list of available open-source implementations. It is important to notice that, given their popularity and great availability of pre-trained models, most of the surveyed papers focus on the distribution of Convolutional Neural Networks (CNN) applied to the computer vision field for tasks like image classification, segmentation or object tracking. But the general aspects and methods presented in these papers can easily be applied to the distribution of other kind of architectures including, but not limited to, Recurrent Neural Networks (RNN) or Transformer Networks.


The remainder of this paper is organized as follows: Section \ref{section:fundamentals} provides an overview of what we mean by distributed inference. Section \ref{section:methodology} presents the survey methodology and provides information about the searched databases, keywords used, and exclusion and inclusion criteria. In Section \ref{section:body}, we analyse each aspect of this problem, proving insights into the current trends and challenges on the field, and promising future research directions. Finally, section \ref{section:conclusions} summarizes the conclusions of this study.

{\footnotesize
\begin{table}[htbp]
	\caption{Glossary of abbreviations used in this survey}
	\begin{center}
		\begin{tabular}{cc} \toprule
			Abbreviation & Meaning\\
			\midrule
			DNN & Deep Neural Network \\
			CNN & Convolutional Neural Network \\
			SNN & Spiking Neural Network \\
			MCU & Microcontroller \\
			GPU & Graphical Processing Unit \\
			FPGA & Field Programmable Gate Array \\
			RPi & Raspberry Pi \\
			\ifm & Input Feature Map \\
			\ofm & Output Feature Map \\
			DAG & Directed Acyclic Graph \\
			GOP & Giga ($10^9$) Operations \\
			GOP/s & Giga ($10^9$) Operations per second\\
			BW & Bandwidth \\
			RL & Reinforcement Learning \\
			FL & Federated Learning \\
			CPU & Central Processing Unit \\
			GPU & Graphical Processing Unit \\
			TPU & Tensor Processing Unit \\
			DSP & Digital Signal Processor \\
			KD & Knowledge Distillation \\
			AE & Auto-encoder \\
			\bottomrule
		\end{tabular}
		\label{tab:glossary}
	\end{center}
\end{table}
}

\section{Fundamentals of the distribution of DNN}
\label{section:fundamentals}

\begin{figure}[!t]
	\centering
	\begin{subfigure}[t]{0.32\textwidth}
		\centering
		\includegraphics[width=\textwidth]{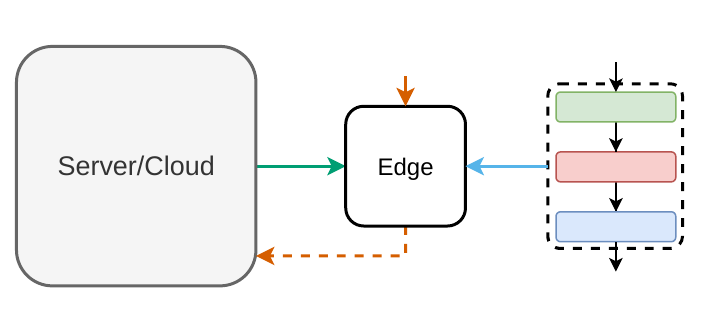}
		\caption{No distribution}
		\label{fig:ml_distribution_0}
	\end{subfigure}
	\begin{subfigure}[t]{0.32\textwidth}
		\centering
		\includegraphics[width=\textwidth]{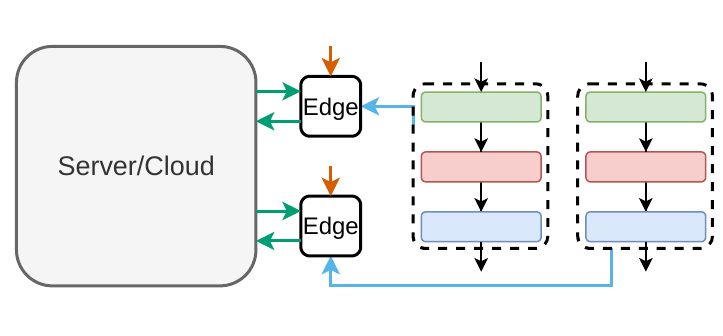}
		\caption{Federated Learning}
		\label{fig:ml_distribution_1}
	\end{subfigure}
	\begin{subfigure}[t]{0.32\textwidth}
		\centering
		\includegraphics[width=\textwidth]{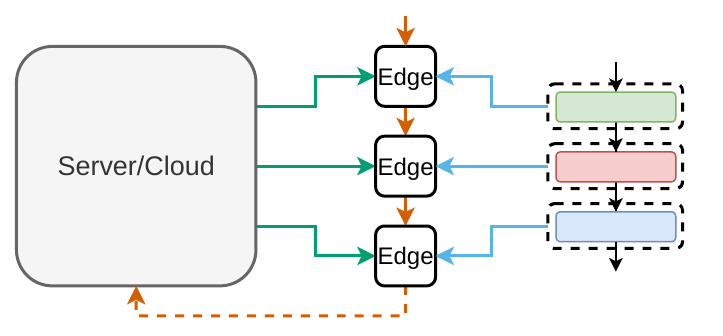}
		\caption{Horizontal distribution}
		\label{fig:ml_distribution_2}
	\end{subfigure}
	\begin{subfigure}[t]{0.32\textwidth}
		\centering
		\includegraphics[width=\textwidth]{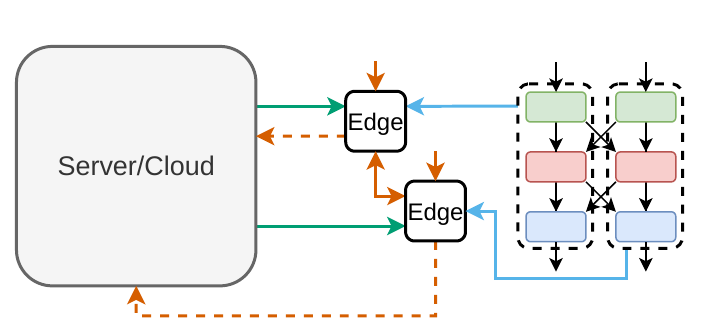}
		\caption{Vertical distribution}
		\label{fig:ml_distribution_3}
	\end{subfigure}
	\begin{subfigure}[t]{0.32\textwidth}
		\centering
		\includegraphics[width=\textwidth]{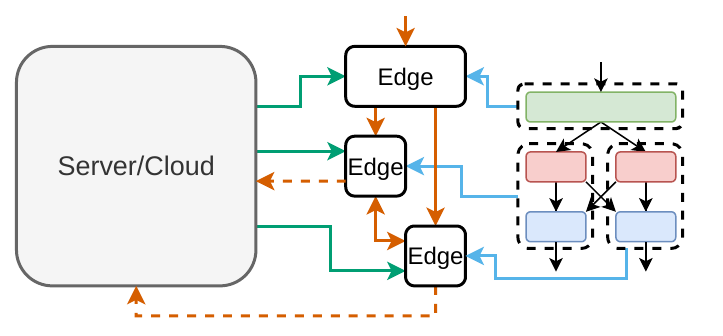}
		\caption{Hybrid distribution}
		\label{fig:ml_distribution_4}
	\end{subfigure}
	\begin{subfigure}[t]{0.32\textwidth}
		\centering
		\includegraphics[width=\textwidth]{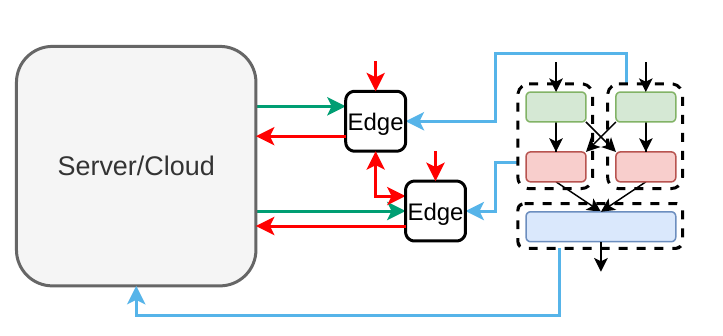}
		\caption{Hybrid distribution including cloud inference}
		\label{fig:ml_distribution_5}
	\end{subfigure}
	\caption{Different options to distribute the inference of a DNN across multiple devices. Green arrows represent configuration parameters. Orange arrows the movement of data between devices. Blue arrows the allocation of layers to particular devices.}
	\label{fig:ml_distribution}
\end{figure}

DNN are already state-of-the-art (SotA) solutions for a variety of tasks, especially in the computer vision field. These algorithms have proven to be well suited to solve a wide range of problems like image classification, segmentation, object detection, tracking, multisensor fusion, etc. For the purpose of this study, a DNN can be described as a set of interconnected layers forming a Direct Acyclic Graph (DAG) $\mathcal{G} = \{\mathcal{V},\mathcal{E}\}$. Each vertex $v \in \mathcal{V}$ equals a particular layer of the model and each edge $(v_i,v_j) \in \mathcal{E}$ represents data dependencies between layers (each vertex can have multiple edges that originate from it or end at it). The complete network can have multiple inputs (although the usual DNN for computer vision normally has only one input: the image to be processed) and multiple output layers (object detection outputs, classification, etc). In a general description, each layer receives one or more tensors, called input feature maps (\ifm). The layer then transforms it, and generates a new tensor, an output feature map (\ofm). Some layers such as convolution or fully connected layers, also have weight and bias tensors associated with the layer, which are used to transform the \ifm. These weights and biases need to be stored in memory, which can be also problematic for small embedded devices given memory constraints. Other layers do not have associated weights but are useful for other purposes, such as concatenation, pointwise addition, pooling, and activation layers.

The DNN's lifetime can be separated in two distinct phases. During the training phase, the weights of the DNN are modified to satisfy a particular target of the application. Because it is beyond the scope of this survey to discuss the distribution of the training phase, we refer the interested reader to other surveys on this topic \cite{baccourPervasiveAIIoT2022}. During the inference phase, an input is inserted into the network and propagated through all the layers until the output is obtained. This phase is characterized by two features of the network: the memory footprint (which depends on the sizes of the weights matrices and the intermediate feature maps between layers) and the GOP required to obtain the output (which depends on the complexity of the selected layers and the DNN architecture, and the input data size). These are two aspects that can prevent DNN from being executed on only one device. If the available memory of the device is insufficient to fit the entire network, it becomes impossible to execute. However, even when the memory is sufficient, if the computing capabilities of a device are limited, it can take a prohibitive amount of time to run just one inference pass of the DNN. The throughput of the system can also be limited, as a single device can not normally start processing a new image until the last image is completely processed.

This is why the concept of distributed inference is important. By distributing the inference pass of a DNN across multiple devices, less weight is stored on each device (so the memory constraint can be met more easily) and the execution can be accelerated by parallelizing computations, thus meeting the latency constraints for a particular application. Of course, memory and latency are not the only possible reasons one could want to distribute the inference pass of a DNN: one could offload a particular kind of layer to one specific hardware because it can be executed in a more efficient manner, minimize energy consumption, or connect the devices in a sequential manner to improve the throughput of the entire system. Moreover, there can also be application specific reasons to distribute the inference of a DNN. For example, when the data generators are \textit{physically} apart from each other. This may be the case in industry or autonomous car scenarios where several sensors located far away must be used as input for a DNN that will make a decision based on the information provided by each sensor. In this case, distributing the first layers of the DNN so that they can be executed near each sensor can help reduce the amount of communicated data to the central system, by preprocessing the raw data from each sensor. The central system would then execute the rest of the DNN and provides the final output.

Figure \ref{fig:ml_distribution} present different approaches to the distribution of DNN, and including the Federated Learning (FL) case, in order to show the difference between it and distributed inference. Figure \ref{fig:ml_distribution_0} shows the basic setup without distribution. The DNN is trained in a powerful server or on the cloud, and the model is downloaded to an edge device which receives the input data and executes the entire DNN (optionally, the output of the DNN can be send to the server/cloud). In FL (Figure \ref{fig:ml_distribution_1}), the DNN inference is not partitioned and each device runs the entirety of the DNN. Each edge device also updates locally the parameters of its model during inference, and regularly send these parameters to the server/cloud where they are merged together with the global model. This improves the global model and at the same time protects the privacy of the raw data processed by each edge device. Notice that this update of the local parameters on each edge device is a training procedure, and this is mostly what separates this case from the distributed inference case. We refer the reader to \cite{baccourPervasiveAIIoT2022} for a survey on distributed methods for FL.

Figures \ref{fig:ml_distribution_2}, \ref{fig:ml_distribution_3}, \ref{fig:ml_distribution_4} and \ref{fig:ml_distribution_5} present different configurations used to run distributed inference. Figure \ref{fig:ml_distribution_2} shows the horizontal distribution case, where each layer (or sequential group of layers) is assigned to one edge device. In this case, the DNN is partitioned in a coarse manner. However, Figure \ref{fig:ml_distribution_3} shows the vertical partitioning scenario where each layer is partitioned in smaller layers and assigned to different devices to improve parallelism. Figures \ref{fig:ml_distribution_4} and \ref{fig:ml_distribution_5} present cases with hybrid distribution configurations, including one where the last layer is assigned to be executed in the server/cloud device.

As can be seen from the multiple examples provided in Figure \ref{fig:ml_distribution}, the complexity of the possible configurations that can be used to distribute the DNN inference pass brings new problems. When designing a distributed system, the following questions naturally arise:

\begin{itemize}
	\item \textbf{Distribution selection:} how can one select an optimized distribution configuration? Where do one partition the network and decide which device to allocate to each partition? Is the search space sufficiently small to try all possible distribution configurations and select the best one? Alternatively, an algorithm needs to be developed to guide the selection of the optimal configuration. If this is the case, how is the problem modelled?
	\item \textbf{Devices:} how many are there available? Are they all equal or do they have different characteristics and features? Are models available to predict the performance (latency and/or energy) of each of them to guide the algorithm's selection, or is profiling needed?
	\item \textbf{Metrics and constraints:} what is the metric that needs to be optimized? Are there more than one, and if that is the case, are there priorities between the metrics? Are there any particular hard constraints that need to be considered? 
	\item \textbf{Adaptability}: does the system need to be adaptable to specific situation-dependent environment changes (BW between devices changes, devices are added or are taken out of the system, input arrival rate increases, etc), or does the distribution configuration need to be calculated only at compile time, and then never change?
\end{itemize}

Inspired by the previous questions, Figure \ref{fig:taxonomy} presents the categorization of the embedded distributed DNN inference papers proposed in this paper. With this, we aim to find common features that would allow researchers to quantitatively and qualitatively compare different studies. In Section \ref{section:body}, each of these categories is analyzed for all the surveyed papers, and challenges and proposals are discussed.

\begin{figure}[!t]
	\centering
	\includegraphics[width=\textwidth]{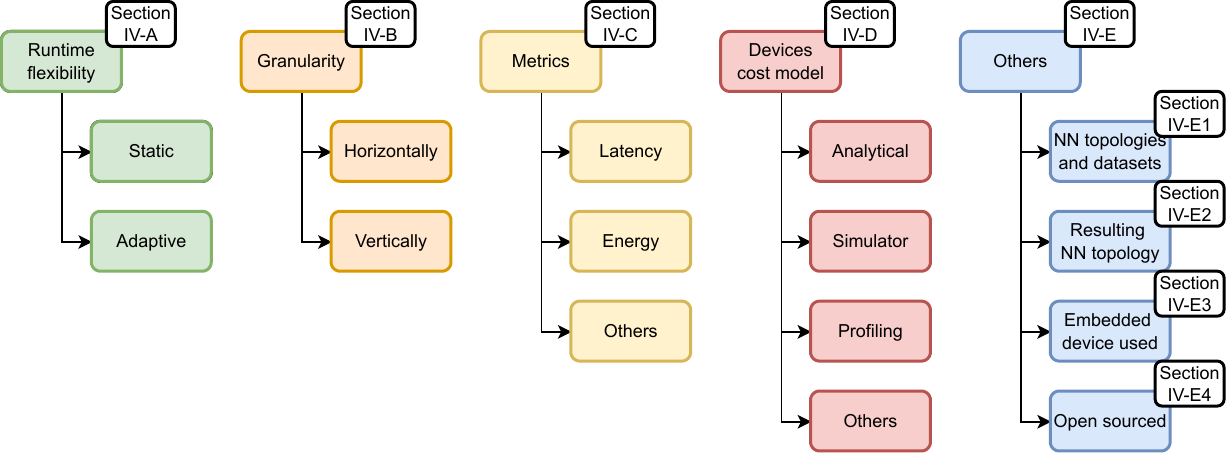}
	\Description{An image showing different boxes which describe the different categorizations of embedded distributed papers that are proposed in this survey, including references to the corresponding section where it is described.}
	\caption{Categorization of embedded distributed DNN inference papers and its position in the review structure}
	\label{fig:taxonomy}
\end{figure} 

\section{Study methodology}
\label{section:methodology}

\begin{figure}[!t]
	\centering
	\includegraphics[width=\textwidth]{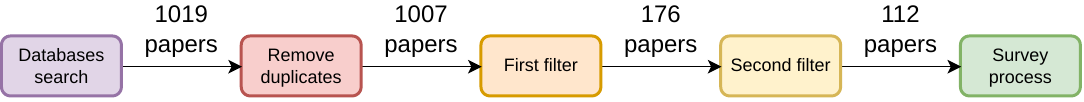}
	\Description{A chain of boxes describing the different steps taken during our systematic review to obtain the final list of papers to survey.}
	\caption{Systematic review process}
	\label{fig:methodology_process}
\end{figure} 

This survey builds on \cite{baccourPervasiveAIIoT2022,chenDistributedLearningWireless2021} but approaches the problem using a systematic review methodology. As such, we first defined the scope of this review as the algorithm and decision methodology for partitioning the neural network. Our aim is to focus on how the allocation and partition problem is modelled and solved, how the features of the available devices and the deployment environment are characterized, what are the most common optimization variables and constraints for these algorithms, and which are the most used metrics to compare them. We also detect gaps and future research opportunities in each of these aspects and propose improvements to the current comparison methodology between distributed inference papers.

{\footnotesize
\begin{table}[htbp]
	\caption{Exclusion and inclusion criteria}\centering
	\begin{tabularx}{\textwidth}{XX} \toprule
		Exclusion criteria & Inclusion criteria \\
		\midrule
		No reviews or surveys are included & Only papers published from 2017 to 2022 \\
		&\\
		No papers on distributed training & Only papers on distributed inference of DNN\\
		&\\
		No papers distributing across server-size GPUs & Only papers that contain embedded devices (MCUs, embedded GPUs, FPGAs, RPis, etc)\\
		&\\
		No papers distributing the inference of Spiking Neural Networks (SNN) & \\
		\bottomrule
	\end{tabularx}
	\label{tab:exclusion_inclusion_criteria}
\end{table}
}

To apply the systematic review methodology, we defined the inclusion and exclusion criteria as shown in Table \ref{tab:exclusion_inclusion_criteria}. To make sure that we are concentrating on SotA publications, we propose to only examine papers from the last six years. The choice to begin the review with papers released in 2017 was made for two reasons. First off, as shown in Figure \ref{fig:work_per_year} this is the year when interest in this field first started to grow up. Second, we wanted to incorporate publications from 2017 like MoDNN \cite{maoMoDNNLocalDistributed2017a} and Neurosurgeon \cite{kangNeurosurgeonCollaborativeIntelligence2017a} that serve as the foundation for many SotA works. Additionally, all papers presenting techniques to distribute the training of DNN are excluded from this survey. We also focus only on papers targeting embedded devices, including but not limited to RPi-type boards, mobile devices, MCUs, embedded GPUs, etc. (papers that distribute the execution of an DNN across an embedded device and the cloud or a more powerful device are also included). 

We selected four databases of peer-reviewed scientific papers to search for: \href{https://ieeexplore.ieee.org/Xplore/home.jsp}{IEEEXplore}, \href{https://dl.acm.org/}{ACM Digital Library}, \href{https://www.sciencedirect.com/}{ScienceDirect} and \href{https://link.springer.com/}{Springer}. To build the search strings for each database, we selected several keywords, as presented in Table \ref{tab:keywords}. When building the search strings, keywords in the same row were combined using an OR operator, and the rows were combined using an AND operator.

{\footnotesize
\begin{table}[htbp]
	\caption{Examples of keywords selected to build the databases search strings}\centering
	\begin{tabularx}{\textwidth}{X} \toprule
		Keywords \\
		\midrule
		Distributed, distribution, distribute, partition, partitioning, partitioned, split, splitting, splitted, cooperative, collaborative\\
		\\
		Inference, coinference, co-inference, prediction, predicted\\
		\\
		Neural network, deep learning, deep neural network, DNN, convolution, convolutional, CNN\\
		\\
		Edge, embedded, accelerator, IOT, Internet of things, FPGA\\
		\bottomrule
	\end{tabularx}
	\label{tab:keywords}
\end{table}
}

Figure \ref{fig:methodology_process} presents a diagram of the methodology process. The first database search, using the selected keywords returned 1019 papers on this topic. After automatic duplicate removal, 1007 papers remained. Next, the first filter was applied using the inclusion and exclusion criteria on the title and abstract of each paper. This reduced the number of studies to 176. Then, a second filter was applied by reading each paper thoroughly, reducing the number of papers to 112. These are the papers reviewed in Section \ref{section:body}.

\section{Analysis of distributed inference papers}
\label{section:body}

As mentioned in Section \ref{section:fundamentals}, the task of distributing a DNN across multiple devices introduces a new set of problems and questions. Different techniques and methods to address each of them can be found in the literature. In this section, we analyze the surveyed papers according to the proposed categories presented in Figure \ref{fig:taxonomy} to identify the strengths and gaps of state-of-the-art implementations.

\subsection{Runtime flexibility}
\label{section:flexibility}

One of the first categorizations that needs to be analysed is the runtime flexibility of the resulting system because it greatly modifies not only the decisions taken by the distribution algorithm, but also what these decisions are. Although both types of algorithms select a metric they want to optimize (and perhaps even some constraints that need to be satisfied), we can differentiate between \textit{static} techniques, where the partitioning and allocation are decided offline and do not change during the lifetime of the system, and \textit{adaptive} techniques, which recalculate their decisions according to the changes observed in their runtime environment and adapt to it to fulfil the selected requirements.

\subsubsection{Static}

52 \% of the reviewed papers belong to the static category. We have included in this category papers that, although claiming to be adaptive, use the word to describe that their algorithm \textit{can be adapted} to optimize different metrics, or that it provides different solutions depending on the available bandwidth but does not adapt to the runtime environment in a dynamic manner.

In the case of static runtime, the distribution algorithm is executed only once, as part of the compilation process of the DNN. Once this is completed, different parts are assigned to each device, and the system is put in operation. Because the distribution is only analyzed during the compile time, this allows the implementation of more complex algorithms, which can run for a long time without hurting the actual operation of the system.


\subsubsection{Adaptive}

More interesting are the papers that can be categorized as \textit{adaptive}, comprising 48 \% of the reviewed papers. They can be further separated according to the most important aspect that defines these kind of papers: the environmental variable that is being observed. As such, we find studies that focus on adapting to changes in the bandwidth between devices, the arrival rate of their inputs, the battery level of the edge device, etc.

The most commonly observed variable is the bandwidth between the devices. Usually, these papers monitor the bandwidth between devices (for example, using the \textit{iPerf} tool \cite{iperf}) and change their allocation decision accordingly to try to minimize or maximize metrics such as latency, Quality of Service (QoS), or load balancing. For example, Autodidactic Neurosurgeon \cite{zhangAutodidacticNeurosurgeonCollaborative2021a} monitors the bandwidth between the edge and the cloud and changes the partition point. In their video processing example, they reported that their system requires approximately 20–80 frames to reach a new stable distribution. Although this is the most common use case in the reviewed adaptive papers, some of them propose different interesting observable variables or methodologies.

AutoScale \cite{kimAutoScaleEnergyEfficiency2020a} uses a Reinforcement Learning (RL) approach to monitor the wireless signal strength between an edge device and the cloud instead of measuring the bandwidth directly. CAMDNN \cite{heidariCAMDNNContentAwareMapping2022} first uses an object detector to generate classification inference tasks which are then reallocated by a scheduler which runs every 5 seconds. \cite{luoCloudEdgeCollaborativeIntelligent2021} also uses RL to observe the state of the environment, but in this case, the state contains the batch size of the requests that are arriving at the distributed system and the current communication channel capacity, both normalized. \cite{yunCooperativeInferenceDNNs2022a} proposes to use Knowledge Distillation (KD) \cite{hinton2015distilling} to train a small network (the student) based on the original network (the teacher), and then dynamically select between the student and the teacher network according to the delay constraint in an IoT device/edge server scenario. \cite{wangDynamicResourceAllocation2021a} proposes a scenario where a DNN is distributed across multiple vehicles (for example, parked cars in a smart city infrastructure), so the distribution algorithm needs to adapt to the number of available vehicles and their computing resources. \cite{SplitComputingDNN2022} explores the optimization of the energy consumption of the distributed system by monitoring the battery level of the IoT device, in order to dynamically decide if it needs to assign its task to a neighbour edge device in order to save battery. \cite{zhaoDeepThingsDistributedAdaptive2018a} proposes a stealing mechanism to distribute workloads across a system of IoT clusters, allowing idle devices to overtake tasks assigned to other devices to improve inference processing.

In this category, the reconfiguration time of the system is extremely important because it defines how quickly the algorithm can adapt to environmental changes. This reconfiguration time considers the time taken to run the distribution algorithm and to reconfigure each device to allow it to execute its newly assigned partitions.

{\footnotesize
\begin{table}[htbp]
	\caption{Papers categorized according to its adaptability, showing the selected observable variable and reported reconfiguration time for adaptive papers}
	\begin{center}
		\begin{tabularx}{1\textwidth}{>{\hsize=.1\hsize}X>{\hsize=.4\hsize}X>{\hsize=.3\hsize}X>{\hsize=.3\hsize}X} \toprule
			Category & Papers & Adaptation variable & Reconfiguration time \\
			\midrule
			Static & \cite{wangFastEnergySavingNeural2021,leeSplittableDNNBasedObject2021a,chenAcceleratingDNNInference2021a,jiangAchievingSuperLinearSpeedup2019,miaoAdaptiveDNNPartition2020a,changEfficientDistributedDeep2019a,banitalebi-dehkordiAutoSplitGeneralFramework2021,dagliAxoNNEnergyAwareExecution2022,zouCAPCommunicationAwareAutomated2022,baeCapellaCustomizingPerception2019a,yangCNNPCEndEdgeCloudCollaborative2022,jiaCoDLEfficientCPUGPU2022a,vanishreeCoInAcceleratedCNN2020a,zhangCommunicationComputationEfficientDeviceEdge2021,shaoCommunicationComputationTradeoffResourceConstrained2020,yangCooperativeDistributedDeep2021a,sbaiCutDistilEncode2021a,parthasarathyDEFERDistributedEdge2022a,sahuDENNIDistributedNeural2021a,houDistrEdgeSpeedingConvolutional2022a,liDistributedDeepLearning2022a,huDistributedInferenceDeep2022a,gacoinDistributingDeepNeural2019a,zhangDynamicDNNDecomposition2021a,zhouDynamicPathBased2021a,yangEdgeComputingNetworking2021a,xueEdgeLDLocallyDistributed2020a,jinEnergyAwareWorkloadAllocation2019,xuEOPEfficientOperator2022a,kressHardwareawarePartitioningConvolutional2022a,songIndustrialVisionOptimization2021a,pachecoInferenceTimeOptimization2020a,yangIntelliEyeUAVTracking2019,fangJointArchitectureDesign2022a,heJointDNNPartition2020a,fuJointOptimizationData2021a,choiLegionTailoringGrouped2021a,naveenLowLatencyDeep2021a,haoMultiAgentCollaborativeInference2022,leeNeuralArchitectureSearch2021a,deyOffloadedExecutionDeep2019a,yangOffloadingOptimizationEdge2021a,jeongOptimalPartitioningDistributed2021a,deyPartitioningCNNModels2018a,parasharProcessorPipeliningMethod2020a,liReceptiveFieldbasedSegmentation2022a,fuSplitComputingVideo2022a,fangTeamNetCollaborativeInference2019a,hsuCooperativeConvolutionalNeural2020,camposdeoliveiraPartitioningConvolutionalNeural2018,hadidiDistributedPerceptionCollaborative2018a,zhouAAIoTAcceleratingArtificial2019,maoMoDNNLocalDistributed2017a,changUltraLowLatencyDistributedDeep2019,teerapittayanonDistributedDeepNeural2017,eshratifarJointDNNEfficientTraining2021,koEdgeHostPartitioningDeep2018,wangADDAAdaptiveDistributed2019,jeongIONNIncrementalOffloading2018,matsubaraDistilledSplitDeep2019,shaoBottleNetEndtoEndApproach2020,eshratifarBottleNetDeepLearning2019} & - & - \\
			\midrule
			Adaptive & \cite{samikwaAdaptiveEarlyExit2022a,niuAdaptiveDeviceEdgeCoInference2022,zengBoomerangOnDemandCooperative2019,huDynamicAdaptiveDNN2019a,DynamicSplitComputing2022,almeidaDynODynamicOnloading2022a,liEdgeAIOnDemand2020,liuEEAIEndedgeArchitecture2021a,huFastAccurateStreaming2020a,jiNovelAdaptiveDNN2022a,shiPrivacyAwareEdgeComputing2019,renEdgeAssistedDistributedDNN2020,huangCLIOEnablingAutomatic2020,baccourDistPrivacyPrivacyAwareDistributed2020} & Bandwidth & -\\
			& \cite{zhangElfAccelerateHighResolution2021,huEnablePipelineProcessing2021a,liuLoADPartLoadAwareDynamic2022} & Bandwidth, computing resources & -\\
			& \cite{zhangAutodidacticNeurosurgeonCollaborative2021a} & Bandwidth & Between 20 and 80 frames\\
			& \cite{kimAutoScaleEnergyEfficiency2020a} & Bandwidth & RL alg.: 25.4 $\mu s$, Q-table: 7.3 $\mu s$.\\
			& \cite{zengCoEdgeCooperativeDNN2021a} & Bandwidth & 10 ms\\
			& \cite{wuHiTDLHighThroughputDeep2022} & Bandwidth & 36 ms\\
			& \cite{liJALADJointAccuracyAnd2018} & Bandwidth & 1.77 ms\\
			& \cite{jiangJointModelTask2022a} & Bandwidth & 0.4 ms\\
			& \cite{zhangRealTimeCooperativeDeep2020} & Bandwidth & Worst for RPI3: 1.68 s, best: 1.09 s\\
			& \cite{luoCloudEdgeCollaborativeIntelligent2021} & Bandwidth, batch size & -\\
			& \cite{zhouAdaptiveParallelExecution2019a,xueDDPQNEfficientDNN2022a} & Bandwidth, device performance & -\\
			& \cite{SplitComputingDNN2022} & Bandwidth, battery level & -\\
			& \cite{kangNeurosurgeonCollaborativeIntelligence2017a} & Bandwidth, load level & -\\
			& \cite{laskaridisSPINNSynergisticProgressive2020} & Bandwidth, load level & 14 ms\\
			& \cite{xuDeepWearAdaptiveLocal2020} & Bandwidth, battery level, processor load level & 0.49\% to 4.21\% of the inference latency\\
			& \cite{zhangAccelerateDeepLearning2020a} & Load of the queues of each device & -\\
			& \cite{hadidiCollaborativeInferencingDeep2020a} & Queue occupancy and device performance & 25 iterations \\
			& \cite{zhangAdaptiveDistributedConvolutional2020a,zhangDeepSlicingCollaborativeAdaptive2021a} & Device performance & -\\
			& \cite{heidariCAMDNNContentAwareMapping2022} & Inference requests (scene complexity) & Local scheduler $<$ 1 ms, global scheduler $\approx$ 2.2 ms.\\
			& \cite{yunCooperativeInferenceDNNs2022a} & Delay constraint, SNR & -\\
			& \cite{huangDeeParHybridDeviceEdgeCloud2019} & Inference requests & - \\
			& \cite{zhaoDeepThingsDistributedAdaptive2018a} & Load queues & -\\
			& \cite{mohammedDistributedInferenceAcceleration2020a} & Inference requests & -\\
			& \cite{khanDistributedInferenceResourceConstrained2022} & Processing resources and communication channel & -\\
			& \cite{wangDynamicResourceAllocation2021a} & Computing resources & Between 3,6 and 17,5 ms\\
			& \cite{huangEnablingLowLatency2021a,yangEfficientInferenceAdaptively2021a,luResourceEfficientDistributedDeep2022} & Task arrival rate & -\\
			& \cite{liEnablingRealtimeAI2022a} & Task deadline requirements & -\\
			& \cite{kroukaEnergyEfficientModelCompression2021} & Channel condition & -\\
			& \cite{dongJointOptimizationDNN2021a} & Amount of devices & -\\
			& \cite{baccourRLPDNNReinforcementLearning2021a} & Devices and classification requests & -\\
			& \cite{tuliSplitPlaceAIAugmented2022a} & Task deadline and device current resource usage & Scheduling time of 9.32 s\\
			\bottomrule
		\end{tabularx}
		\label{tab:flexibility}
	\end{center}
\end{table}
}

\subsubsection{Trends and challenges}

Table \ref{tab:flexibility} presents our findings across this categorization of distributed inference papers. As can be seen, the current trend is to focus on static runtime solutions, although a significant amount of surveyed papers focus on adaptive solutions. From these studies, the most used variable to determine how the partitioning must be adapted is the bandwidth of the system. A small number of studies have focused on load-balancing optimization by controlling the number of partitions assigned to each device. An even smaller number of studies have focused on device performance, which can change because of several factors (processor load level, battery level, etc.). Finally, some studies change their distribution depending on the properties of their input data: task arrival rate or deadline requirements.

There are two promising research directions on this particular aspect. The first one would be to explore other adaptation variables, as there is a clear research focus on adapting respecting to bandwidth changes. The second one is the metric used to evaluate these papers. As Table \ref{tab:flexibility} shows the reconfiguration time is reported in different ways (actual time, percentage of the inference latency, iterations of the algorithm, etc.). Founding a uniform metric to be able to quickly compare these papers between them would be an important addition to the field.

\subsection{Granularity of the partition points}
\label{section:granularity}



Now that the adaptability of the distributed system is defined, we can focus on other important aspects of the distribution algorithm: how the DNN is partitioned. This can be done in two different ways, which are presented in this section.

\subsubsection{Horizontally}

As presented in Figure \ref{fig:ml_distribution_2}, \textit{horizontal} splitting (sometimes also called \textit{sequential}) partitions the DNN in a pipelined manner by allocating a group of consecutive layers to the same device. This partitioning method provides a smaller search space with fewer variables to be tuned for the distribution algorithm, thus generating simpler and faster problem formulations. Therefore, partitioning decisions can be obtained more quickly, and algorithms that make these decisions can run on more constrained devices. This is the reason why approximately 80\% of the adaptive papers presented in Section \ref{section:flexibility} use this option as its splitting method.

Because of the sequential nature of the resulting distributed system, this method is typically used when distributing the execution of an DNN to optimize the throughput of the system (see Section \ref{section:target_to_optimize}). This is the case of the distribution across an edge device and cloud, where only one splitting point needs to be selected. However, other studies have also selected this method when distributing across more complex setups, where multiple partitioning points need to be selected, for example, in edge-fog-cloud systems.

\subsubsection{Vertically}
\label{section:vertical_partition}

In contrast, vertical splitting (sometimes also called parallel) partitions the \ofm generated by a layer, creating two or more \textit{sublayers} that can be executed by different devices in parallel, which only needs a portion of the original \ifm (or weight tensor) to successfully execute its calculations. However, as shown in Figure \ref{fig:ml_distribution_3}, this method introduces synchronization points and merging layers to provide the correct \ifm for the next partitioned layer. In DNNs, this phenomenon appears because of the overlapping nature of convolutions, which depends on its kernel sizes and strides, and may require data from multiple sublayers output (refer to \cite{baccourPervasiveAIIoT2022} for a detailed analysis of how this \ofm partitioning modifies the properties of the generated sublayers, including but not limited to memory consumption, transmitted data, and merging strategy). As this method introduces more decision variables for the distribution algorithm and significantly expands the search space, it is usually selected for papers that fall under the static runtime flexibility category.

\begin{figure}[!t]
	\centering
	\includegraphics[width=\textwidth]{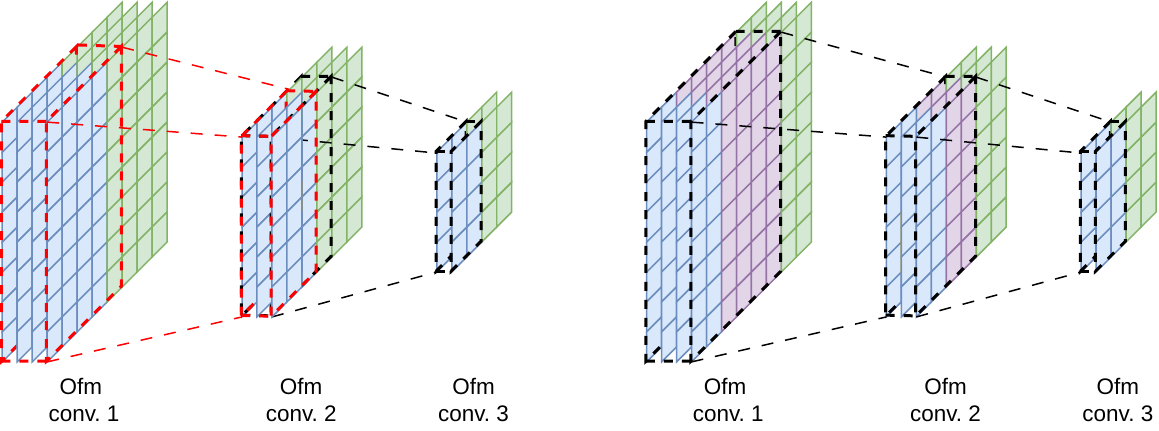}
	\Description{Two ways to assign computation when the network is partitioned using vertical partitioning and two devices are available. Each figure shows 3 feature maps, and dashed lines show the input data dependencies of each output feature map. On the left, the naive partition solution shows how one device needs to fetch the data generated by another device in order to be able to calculate its assigned output feature map. On the right, by sending redundant data to both devices (this redundancy is represented with a different colour in the figure), each one of them can generate multiple output feature maps before needing to fetch data from other devices.}
	\caption{Parallel generation of the \ofm of sequential layers. Left: naive partition, no fusion, device assigned to generate blue \ofm needs data from the one assigned the green one. Right: purple \ofm represents the data that is generated/needed by both devices, in order to avoid data transfer between devices.}
	\label{fig:layer_fusion}
\end{figure} 


In this partitioning method, there are several ways to partition the \ofm and generate the sublayers. First, we can separate between \textit{segment-based} and \textit{grid-based} partitions. In the first option, the original \ofm is partitioned across only one axis (e.g., across its height), which generates \ofm that resemble stripes of the original \ofm. In the second option, the original \ofm is partitioned across two axes (for example, width and height), thus generating a grid of \ofm. \cite{mohammedDistributedInferenceAcceleration2020a} provides a small analysis of these two approaches and concludes that segment-based partitioning is beneficial because it requires the transmission of fewer redundant values than grid-based partitions.

One popular method of partitioning \ofm is to generate equal-sized partitions that are equal to the number of available devices. This is useful when all available devices have the same computing capabilities, because great load balancing can be achieved. However, it is not optimal when there are very dissimilar devices, because very powerful devices can finish their assigned sublayers before the other ones and become idle until the other less powerful devices finish their computations.

To improve upon this idea, studies such as Legion \cite{choiLegionTailoringGrouped2021a} generate partitions that are proportional to the computing capacity of each available device. For each layer, partitions are generated such that devices with more computing power are assigned more demanding sublayers (in terms of operations needed to generate its \ofm), and less powerful devices are assigned smaller sublayers. This can be used to improve the load balancing of the distributed system when devices with heterogeneous computing capabilities are available.

Both of these methods are \textit{fixed} partitioning methods, which means that the number of partitions in each layer is selected before running the distribution algorithm. This is sufficient for some use cases, but does not guarantee that these selected partitions are the best way of partitioning each layer. In the more complex and generic problem formulation, the partition indexes (i.e., the points where the \ofm needs to be split) are dynamically selected when the distribution algorithm is executed. This clearly generates a huge search space for the algorithm but can help find better distributions.

To mitigate the synchronization and merging problem of the vertical partitioning method to some extent, a technique called \textit{layer fusion} can be used. Without layer fusion, data needs to be exchanged between devices before they are ready to calculate their assigned \ofm (left diagram in Figure \ref{fig:layer_fusion}, where it can be seen how a naive partition between two devices forces each of them to query data from the other each time they need to calculate a new \ofm). On the other hand, by using layer fusion, layers are partitioned in such a way that the \ofm produced by the resulting sublayers are \textit{exactly} the \ifm needed by the sublayers of the second layer. Consecutive sublayers are then assigned to the same device, thereby reducing the need to exchange data between them (right diagram in Figure \ref{fig:layer_fusion}). These entire sequences of sublayers (chains of fused sublayers) are executed on the same device without any need to communicate data to other devices. However, this method also has some drawbacks. First, there is an overlap between \ofm generated by the fused sublayers, which generates redundant computations across devices (purple sections in Figure \ref{fig:layer_fusion}). This redundancy becomes more apparent and increases with longer fused chains of sublayers. Although reducing the inter-device communication is usually beneficial for the latency and/or energy of the entire distributed system, it needs to be balanced with the length of these fused chains to prevent redundant computations (and the initial cost of sending redundant data to each device) from impacting the system performance.

Although these are the most commonly used vertical partitioning regimes, there are particular papers that have proposed novel ideas to partition the inference task differently, but should still be considered inside this category because they partition the execution of the DNN in a parallel fashion across multiple devices. For example, Elf \cite{zhangElfAccelerateHighResolution2021} first generated region proposals for the input image, which were then distributed to each available edge server to be processed in parallel. EDDL \cite{changEfficientDistributedDeep2019a} uses both partitioning methodologies. In Coln \cite{vanishreeCoInAcceleratedCNN2020a}, batches are processed by parallel devices to improve throughput, but each device executes the entire DNN. TeamNet \cite{fangTeamNetCollaborativeInference2019a} generates multiple, smaller, \textit{expert models} that are then executed on different edge devices.

{\footnotesize
\begin{table}[htbp]
	\caption{Papers categorized according to the granularity of its distribution}
	\begin{center}
		\begin{tabularx}{\textwidth}{>{\hsize=.15\hsize}X>{\hsize=.15\hsize}X>{\hsize=.7\hsize}X} \toprule
			Category & Subcategory & Papers\\
			\midrule
			Horizontal & Two devices & \cite{leeSplittableDNNBasedObject2021a,samikwaAdaptiveEarlyExit2022a,niuAdaptiveDeviceEdgeCoInference2022,banitalebi-dehkordiAutoSplitGeneralFramework2021,zhangAutodidacticNeurosurgeonCollaborative2021a,zengBoomerangOnDemandCooperative2019,luoCloudEdgeCollaborativeIntelligent2021,yangCNNPCEndEdgeCloudCollaborative2022,zhangCommunicationComputationEfficientDeviceEdge2021,shaoCommunicationComputationTradeoffResourceConstrained2020,yunCooperativeInferenceDNNs2022a,sbaiCutDistilEncode2021a,khanDistributedInferenceResourceConstrained2022,huDynamicAdaptiveDNN2019a,zhouDynamicPathBased2021a,wangDynamicResourceAllocation2021a,DynamicSplitComputing2022,almeidaDynODynamicOnloading2022a,liEdgeAIOnDemand2020,yangEdgeComputingNetworking2021a,liuEEAIEndedgeArchitecture2021a,huEnablePipelineProcessing2021a,jinEnergyAwareWorkloadAllocation2019,kroukaEnergyEfficientModelCompression2021,kressHardwareawarePartitioningConvolutional2022a,wuHiTDLHighThroughputDeep2022,songIndustrialVisionOptimization2021a,pachecoInferenceTimeOptimization2020a,yangIntelliEyeUAVTracking2019,liJALADJointAccuracyAnd2018,jiangJointModelTask2022a,dongJointOptimizationDNN2021a,liuLoADPartLoadAwareDynamic2022,leeNeuralArchitectureSearch2021a,jiNovelAdaptiveDNN2022a,deyOffloadedExecutionDeep2019a,yangOffloadingOptimizationEdge2021a,shiPrivacyAwareEdgeComputing2019,fuSplitComputingVideo2022a,SplitComputingDNN2022,zhangRealTimeCooperativeDeep2020,xuDeepWearAdaptiveLocal2020,renEdgeAssistedDistributedDNN2020,koEdgeHostPartitioningDeep2018,wangADDAAdaptiveDistributed2019,kangNeurosurgeonCollaborativeIntelligence2017a,laskaridisSPINNSynergisticProgressive2020,jeongIONNIncrementalOffloading2018,huangCLIOEnablingAutomatic2020,matsubaraDistilledSplitDeep2019,shaoBottleNetEndtoEndApproach2020,eshratifarBottleNetDeepLearning2019} \\
			& Multiple devices & \cite{kimAutoScaleEnergyEfficiency2020a,zhangAccelerateDeepLearning2020a,chenAcceleratingDNNInference2021a,dagliAxoNNEnergyAwareExecution2022,xueDDPQNEfficientDNN2022a,huangDeeParHybridDeviceEdgeCloud2019,parthasarathyDEFERDistributedEdge2022a,gacoinDistributingDeepNeural2019a,huangEnablingLowLatency2021a,liEnablingRealtimeAI2022a,huFastAccurateStreaming2020a,heJointDNNPartition2020a,haoMultiAgentCollaborativeInference2022,parasharProcessorPipeliningMethod2020a,luResourceEfficientDistributedDeep2022,tuliSplitPlaceAIAugmented2022a,yangEfficientInferenceAdaptively2021a,zhouAAIoTAcceleratingArtificial2019,changUltraLowLatencyDistributedDeep2019,eshratifarJointDNNEfficientTraining2021} \\
			\hline 
			Vertical & No layer fusion & \cite{wangFastEnergySavingNeural2021,jiangAchievingSuperLinearSpeedup2019,zhangAdaptiveDistributedConvolutional2020a,miaoAdaptiveDNNPartition2020a,changEfficientDistributedDeep2019a,vanishreeCoInAcceleratedCNN2020a,sahuDENNIDistributedNeural2021a,liDistributedDeepLearning2022a,mohammedDistributedInferenceAcceleration2020a,huDistributedInferenceDeep2022a,zhangElfAccelerateHighResolution2021,xuEOPEfficientOperator2022a,jeongOptimalPartitioningDistributed2021a,deyPartitioningCNNModels2018a,baccourRLPDNNReinforcementLearning2021a,fangTeamNetCollaborativeInference2019a,baccourDistPrivacyPrivacyAwareDistributed2020,hsuCooperativeConvolutionalNeural2020,camposdeoliveiraPartitioningConvolutionalNeural2018,hadidiDistributedPerceptionCollaborative2018a,hadidiCollaborativeInferencingDeep2020a,maoMoDNNLocalDistributed2017a,teerapittayanonDistributedDeepNeural2017} \\
			& Layer fusion & \cite{zhouAdaptiveParallelExecution2019a,zouCAPCommunicationAwareAutomated2022,baeCapellaCustomizingPerception2019a,jiaCoDLEfficientCPUGPU2022a,zengCoEdgeCooperativeDNN2021a,yangCooperativeDistributedDeep2021a,zhangDeepSlicingCollaborativeAdaptive2021a,zhaoDeepThingsDistributedAdaptive2018a,houDistrEdgeSpeedingConvolutional2022a,zhangDynamicDNNDecomposition2021a,xueEdgeLDLocallyDistributed2020a,fangJointArchitectureDesign2022a,fuJointOptimizationData2021a,choiLegionTailoringGrouped2021a,naveenLowLatencyDeep2021a,liReceptiveFieldbasedSegmentation2022a} \\
			\bottomrule
		\end{tabularx}
		\label{tab:granularity}
	\end{center}
\end{table}
}

\subsubsection{Trends and challenges}

From Table \ref{tab:granularity}, it can be seen that there is a clear tendency in current studies to focus on the horizontal distribution of DNN, mostly across two devices (for example, edge-cloud systems). Multiple factors may explain this: its simplicity when compared to vertical partitioning, the necessity of processing the data in the cloud that naturally generates pipeline-type systems, or the need to improve the throughput when compared to edge-only or cloud-only systems. Focusing on generic solutions to deploy DNNs across n-tier systems seems like a promising research direction.

However, this study also shows that there are gaps in distributed inference research, particularly in the vertical partitioning category. For the vertical category, we decided to subcategorize it across techniques using or not using layer fusion (a categorization across two or multiple devices does not make sense in this case, as most vertical partitioning methods are generalized to be applied to \textit{n} devices). Future papers should continue focusing on layer fusion, as only 40\% of the vertical partitioning papers have explore this technique.

\subsection{Metrics}
\label{section:target_to_optimize}


Now that the partitioning and allocation options, together with the flexibility of the resulting distributed system are already defined, we can analyse the different optimization targets and constraints. We are going to call both of them metrics. These are grouped together in this section because what some papers used as optimization targets others use as constraints, and the other way around. We define the metrics that are being minimized (or maximized, depending on the metric) during the execution of the distribution algorithm as \textit{optimization targets}. We define the limit values of the metrics (which can be different from those used as optimization targets) that need to be respected once the algorithm reaches a distribution decision as \textit{constraints}.

Table \ref{tab:target_to_optimize} presents the optimization metrics of the surveyed studies. There are 3 basic metrics that are the most commonly used across these papers. These are \textit{latency}, \textit{throughput} and \textit{energy}. By latency (sometimes called \textit{completion time}, \textit{inference delay}, and others), we refer to the time required for the system to run the entire process from obtaining the input data to generating the output result. This metric not only takes into account the time it takes for each device to execute its assigned layers but also needs to take into account the delays and consequences of the communication channels. By throughput, we refer to the number of inputs that the system can process per second, which is typically calculated as stated in \cite{huDynamicAdaptiveDNN2019a}, where the throughput of the entire system is simply the inverse of the maximum latency across the groups of layers executed on each device. By energy, we refer to the consumed energy per inference, which is equal to the energy needed to communicate and process one input.

However, we also find more particular metrics which, although used only in some of the reviewed papers, provide an interesting example of the methods that can be used to improve the design of a distributed system.

As such, we find that a number of studies use \textit{communicated data} (usually measured in bytes or one of its related units) as one of their metrics. The reason behind this selection is that, if less data is communicated between devices, the three main metrics described previously are also indirectly optimized. Although this seems logical, it makes sense only when the communication channels between devices are much slower than the actual devices. If this is the case, then fusion techniques can be safely used to reduce the amount of communicated data because the overhead produced by the redundant calculations does not significantly affect the three basic metrics. However, if the devices are slower or consume a comparable amount of energy compared with the communication channel, then this overhead becomes similar or sometimes even greater than the time or energy saved by using fusion techniques. This can negatively affect the three main metrics.

Another interesting metric is \textit{accuracy}, which is used in studies that not only distribute the DNN, but also use quantization techniques to reduce the bit width used on edge devices. There are diverse reasons for using these methods. AutoScale \cite{kimAutoScaleEnergyEfficiency2020a} distributes the inference across devices (CPU, GPU and TPU/DSP) where each one supports different quantizations. During runtime, AutoScale chooses which parts of the DNN are executed by each one of them. Because using smaller quantizations can negatively affect the accuracy, AutoScale finds a trade-off between offloading to devices that support working with smaller quantization (which are more efficient) and running on the CPU using a standard floating-point representation. CNNPC \cite{yangCNNPCEndEdgeCloudCollaborative2022} uses quantization to reduce the amount of data transmitted between the devices. On the other hand, Edgent \cite{liEdgeAIOnDemand2020} also uses accuracy as a metric, but does not focus on quantization. Edgent trains a model with multiple exit points, each of which provides different accuracies and complexities in terms of GOP. During runtime, Edgent selects the exit point that maximizes accuracy under a given latency constraint.

Other metrics do not appear that often, but present rich additions to the distributed inference landscape. In contrast to all the other reviewed papers, DENNI \cite{sahuDENNIDistributedNeural2021a} takes a different approach by optimizing the minimum number of devices required to run an DNN while considering the memory constraints of the devices. \cite{liEnablingRealtimeAI2022a} uses an adaptive technique that instead of improving the latency of one particular inference, improves the \textit{utility} of the distributed system: how many inference jobs achieve their particular latency deadline in a given timeslot. Given the subscription models of modern clouds, it makes sense for papers distributing across edge and cloud systems to consider the cost per hour of using the cloud service as part of their optimization metric. This is one of the metrics used in \cite{almeidaDynODynamicOnloading2022a,laskaridisSPINNSynergisticProgressive2020,dongJointOptimizationDNN2021a}. Finally, current user privacy concerns become an issue to consider in these distributed systems, as data is shared across devices. It is important to limit the possibility of reconstructing the original data if a third party intercepts some of these intermediate results. As such, papers like
\cite{shiPrivacyAwareEdgeComputing2019,baccourDistPrivacyPrivacyAwareDistributed2020} provide metrics to measure how much information can be retrieved from each \ofm, and take this into account when partitioning the DNN.

{\footnotesize
\begin{table}[htbp]
	\caption{Papers categorized according to the target they aim to optimize}
	\begin{center}
		\begin{tabularx}{\textwidth}{>{\hsize=.15\hsize}X>{\hsize=.2\hsize}X>{\hsize=.65\hsize}X} \toprule
			\multicolumn{1}{l}{Group}       & Metric                       &  Papers\\
			\midrule
			Time related  & Latency                             &  \cite{wangFastEnergySavingNeural2021,leeSplittableDNNBasedObject2021a,zhangAccelerateDeepLearning2020a,chenAcceleratingDNNInference2021a,jiangAchievingSuperLinearSpeedup2019,zhangAdaptiveDistributedConvolutional2020a,miaoAdaptiveDNNPartition2020a,zhouAdaptiveParallelExecution2019a,changEfficientDistributedDeep2019a,banitalebi-dehkordiAutoSplitGeneralFramework2021,zhangAutodidacticNeurosurgeonCollaborative2021a,dagliAxoNNEnergyAwareExecution2022,heidariCAMDNNContentAwareMapping2022,luoCloudEdgeCollaborativeIntelligent2021,jiaCoDLEfficientCPUGPU2022a,shaoCommunicationComputationTradeoffResourceConstrained2020,yangCooperativeDistributedDeep2021a,huangDeeParHybridDeviceEdgeCloud2019,zhangDeepSlicingCollaborativeAdaptive2021a,zhaoDeepThingsDistributedAdaptive2018a,houDistrEdgeSpeedingConvolutional2022a,mohammedDistributedInferenceAcceleration2020a,khanDistributedInferenceResourceConstrained2022,huDistributedInferenceDeep2022a,zhangDynamicDNNDecomposition2021a,zhouDynamicPathBased2021a,wangDynamicResourceAllocation2021a,DynamicSplitComputing2022,xueEdgeLDLocallyDistributed2020a,liuEEAIEndedgeArchitecture2021a,huEnablePipelineProcessing2021a,huangEnablingLowLatency2021a,xuEOPEfficientOperator2022a,songIndustrialVisionOptimization2021a,pachecoInferenceTimeOptimization2020a,yangIntelliEyeUAVTracking2019,liJALADJointAccuracyAnd2018,fangJointArchitectureDesign2022a,heJointDNNPartition2020a,jiangJointModelTask2022a,fuJointOptimizationData2021a,choiLegionTailoringGrouped2021a,liuLoADPartLoadAwareDynamic2022,leeNeuralArchitectureSearch2021a,jiNovelAdaptiveDNN2022a,deyPartitioningCNNModels2018a,liReceptiveFieldbasedSegmentation2022a,luResourceEfficientDistributedDeep2022,baccourRLPDNNReinforcementLearning2021a,SplitComputingDNN2022,fangTeamNetCollaborativeInference2019a,zhangRealTimeCooperativeDeep2020,baccourDistPrivacyPrivacyAwareDistributed2020,hsuCooperativeConvolutionalNeural2020,hadidiCollaborativeInferencingDeep2020a,zhouAAIoTAcceleratingArtificial2019,changUltraLowLatencyDistributedDeep2019,wangADDAAdaptiveDistributed2019,jeongIONNIncrementalOffloading2018,huangCLIOEnablingAutomatic2020} \\
			& Throughput                          &  \cite{baeCapellaCustomizingPerception2019a,vanishreeCoInAcceleratedCNN2020a,parthasarathyDEFERDistributedEdge2022a,liDistributedDeepLearning2022a,huDynamicAdaptiveDNN2019a,zhangElfAccelerateHighResolution2021,huFastAccurateStreaming2020a,wuHiTDLHighThroughputDeep2022,parasharProcessorPipeliningMethod2020a,yangEfficientInferenceAdaptively2021a,hadidiDistributedPerceptionCollaborative2018a} \\
			& Latency, accuracy                   & \cite{niuAdaptiveDeviceEdgeCoInference2022,zengBoomerangOnDemandCooperative2019,yangCNNPCEndEdgeCloudCollaborative2022,yunCooperativeInferenceDNNs2022a,gacoinDistributingDeepNeural2019a,yangEdgeComputingNetworking2021a} \\
			& Latency, communication              & \cite{naveenLowLatencyDeep2021a,maoMoDNNLocalDistributed2017a,matsubaraDistilledSplitDeep2019} \\
			& Latency, privacy                    & \cite{shiPrivacyAwareEdgeComputing2019} \\
			& Latency, cost                                & \cite{dongJointOptimizationDNN2021a} \\
			& Latency, throughput, cost, accuracy & \cite{almeidaDynODynamicOnloading2022a,laskaridisSPINNSynergisticProgressive2020} \\
			\midrule
			Energy related & Energy                              & \cite{zengCoEdgeCooperativeDNN2021a,jinEnergyAwareWorkloadAllocation2019} \\
			& Energy, accuracy                    & \cite{kroukaEnergyEfficientModelCompression2021} \\
			\midrule
			Time and energy related           & Latency, energy                     & \cite{samikwaAdaptiveEarlyExit2022a,kressHardwareawarePartitioningConvolutional2022a,haoMultiAgentCollaborativeInference2022,deyOffloadedExecutionDeep2019a,yangOffloadingOptimizationEdge2021a,tuliSplitPlaceAIAugmented2022a,xuDeepWearAdaptiveLocal2020,renEdgeAssistedDistributedDNN2020,eshratifarJointDNNEfficientTraining2021,kangNeurosurgeonCollaborativeIntelligence2017a} \\
			& Energy,  accuracy, throughput       & \cite{koEdgeHostPartitioningDeep2018} \\
			& Latency, energy, cost               & \cite{xueDDPQNEfficientDNN2022a} \\
			\midrule
			Others         & Accuracy                            & \cite{liEdgeAIOnDemand2020,jeongOptimalPartitioningDistributed2021a,fuSplitComputingVideo2022a} \\
			& Accuracy, memory                    & \cite{sbaiCutDistilEncode2021a} \\
			& Accuracy, sparsity                  & \cite{zhangCommunicationComputationEfficientDeviceEdge2021} \\
			& Communication                       & \cite{camposdeoliveiraPartitioningConvolutionalNeural2018,teerapittayanonDistributedDeepNeural2017} \\
			& Communication, accuracy             & \cite{zouCAPCommunicationAwareAutomated2022,shaoBottleNetEndtoEndApproach2020,eshratifarBottleNetDeepLearning2019} \\
			& Utility                             & \cite{liEnablingRealtimeAI2022a} \\
			& Devices                             & \cite{sahuDENNIDistributedNeural2021a} \\
			\bottomrule
		\end{tabularx}
		\label{tab:target_to_optimize}
	\end{center}
\end{table}
}

When discussing metrics, it is also important to mention the most common measurements used to \textit{compare} distributed inference between them. Although absolute values can be used for comparison purposes (paper A achieved a latency of X ms for network N, surpassing B, who only achieved a latency of Y ms on the same network), this is highly dependent on the capabilities of the available devices, the communication channel parameters (including its model for papers reporting analytical or simulation results), software stack used to compile the partitions of the DNN (for studies reporting actual measurements), DNN configurations (input size, feature map and weight quantization, model), and more factors. To compare these absolute values, one should replicate the test setup by considering all these characteristics. However, because most studies do not provide all this information, it becomes nearly impossible to do so.

As such, \textit{relative} values are normally used to compare the studies between them. For example, studies that partition the neural network in a horizontal manner to distribute its execution across an edge-cloud system tend to report the relative speedup (or metric improvement) against an \textit{edge-only} setup and against a \textit{cloud-only} setup. By doing so, they are able to demonstrate the convenience of using distributed inference versus a \textit{single-device} setup. Vertical partitioning studies tend to report the relative speedup (or metric improvement) per device added. As such, they show how much the optimized metric can be improved by adding more devices in parallel. Because adding more devices helps parallelize the execution of the DNN, but also increases the redundant data that need to be exchanged between devices, papers that use vertical partitioning normally find the Pareto-optimal number of devices for a given distributed system. 

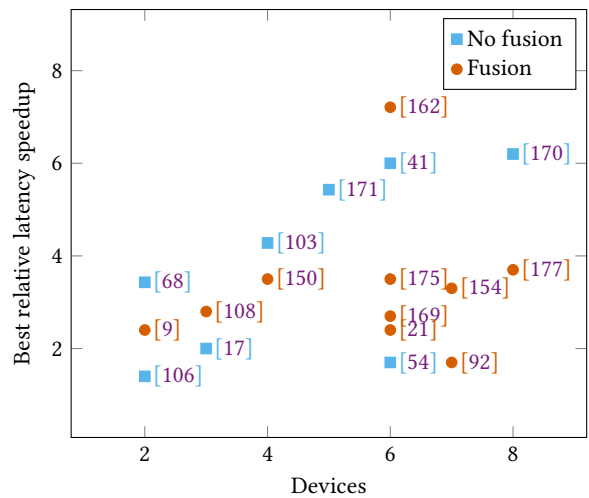
\begin{figure}[htbp]
	\centering
	\begin{tikzpicture}
		\begin{axis}[
			enlargelimits=0.2,
			xlabel=Devices,
			ylabel=Best relative latency speedup,
			ymax=8,
			xmax=8,
			every node near coord/.append style={anchor=west},
			legend cell align={left}
			]
			\addplot[nodes near coords,only marks,
			point meta=explicit symbolic, mark=square*,mark options={my_blue}]
			table[meta=label] {
				x y label
				2 3.43 \cite{jiangAchievingSuperLinearSpeedup2019}
				8 6.2 \cite{zhangAdaptiveDistributedConvolutional2020a}
				2 1.4 \cite{miaoAdaptiveDNNPartition2020a}
				3 2 \cite{changEfficientDistributedDeep2019a}
				6 1.7 \cite{huDistributedInferenceDeep2022a}
				5 5.43 \cite{zhangElfAccelerateHighResolution2021}
				6 6 \cite{hadidiCollaborativeInferencingDeep2020a}
				4 4.28 \cite{maoMoDNNLocalDistributed2017a}
			};
			\addplot[nodes near coords,only marks,
			point meta=explicit symbolic, mark=*,mark options={my_red}]
			table[meta=label] {
				x y label
				8 3.7 \cite{zhouAdaptiveParallelExecution2019a}
				2 2.4 \cite{baeCapellaCustomizingPerception2019a}
				6 7.21 \cite{zengCoEdgeCooperativeDNN2021a}
				7 3.3 \cite{yangCooperativeDistributedDeep2021a}
				6 2.7 \cite{zhangDeepSlicingCollaborativeAdaptive2021a}
				6 3.5 \cite{zhaoDeepThingsDistributedAdaptive2018a}
				4 3.5 \cite{xueEdgeLDLocallyDistributed2020a}
				6 2.4 \cite{choiLegionTailoringGrouped2021a}
				3 2.8 \cite{naveenLowLatencyDeep2021a}
				7 1.7 \cite{liReceptiveFieldbasedSegmentation2022a}
			};
			\legend{No fusion, Fusion}
		\end{axis}
	\end{tikzpicture}
	\Description{A scatter plot showing the best relative latency speedup reported by each vertical partitioning  paper against the number of devices used to achieve said speedup. The corresponding reference is provided for each point in the plot. A different colour is used to group papers where fusion is used.}
	\caption{Best latency speedup (when compared against the case with only one edge device) reported for papers using vertical partitioning}
	\label{fig:vertical_speedup}
\end{figure}

\begin{figure}[!ht]
	\centering
	\begin{tikzpicture}
		\begin{axis}[
			enlargelimits=0.2,
			xlabel=Devices,
			ylabel=Best relative energy savings,
			legend cell align={left}
			]
			\addplot[nodes near coords,only marks,
			point meta=explicit symbolic, mark=*,mark options={my_blue}]
			table[meta=label] {
				x y label
				6 1.6 \cite{zengCoEdgeCooperativeDNN2021a}
			};
			\addplot[nodes near coords,only marks,
			point meta=explicit symbolic, mark=square*,mark options={my_red}]
			table[meta=label] {
				x y label
				3 9.8 \cite{kimAutoScaleEnergyEfficiency2020a}
				2 6.9 \cite{xuDeepWearAdaptiveLocal2020}
				2 2.3 \cite{koEdgeHostPartitioningDeep2018}
			};
			\legend{Vertical, Horizontal}
		\end{axis}
	\end{tikzpicture}
	\Description{A scatter plot showing the best relative energy savings reported by each paper against the number of devices used to achieve said savings. The corresponding reference is provided for each point in the plot. A different colour is used to group papers according to the partitioning scheme used.}
	\caption{Best energy savings (when compared against the case with only one edge device) reported.}
	\label{fig:energy_savings}
\end{figure}
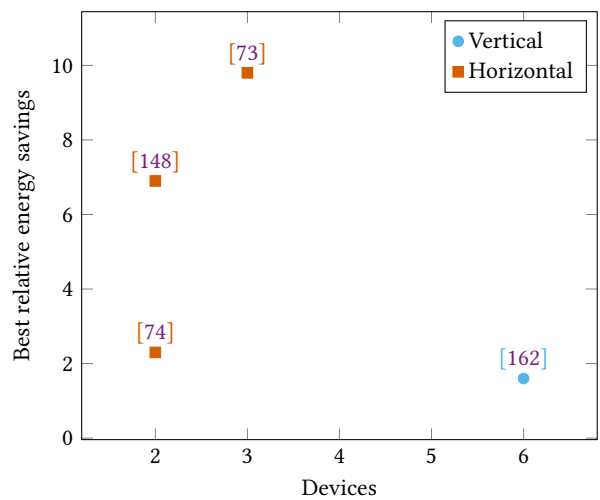

\begin{figure}[!ht]
	\centering
	\begin{tikzpicture}
		\begin{axis}[
			width=\textwidth,
			height=0.3\textwidth,
			xlabel=Papers,
			ylabel=Best relative \\ latency speedup,
			ylabel style={align=center},
			ybar,
			ymax=35,
			ymin=0,
			bar shift=0pt,
			xtick={1,2,3,4,5,6,7,8,9,10,11,12,13,14,15},
			xticklabels={
				\cite{liuLoADPartLoadAwareDynamic2022},
				\cite{xuDeepWearAdaptiveLocal2020},
				\cite{chenAcceleratingDNNInference2021a},
				\cite{zhouDynamicPathBased2021a},
				\cite{heidariCAMDNNContentAwareMapping2022},
				\cite{wangADDAAdaptiveDistributed2019},
				\cite{yangCNNPCEndEdgeCloudCollaborative2022},
				\cite{huDynamicAdaptiveDNN2019a},
				\cite{zhangRealTimeCooperativeDeep2020},
				\cite{parasharProcessorPipeliningMethod2020a},
				\cite{liuEEAIEndedgeArchitecture2021a},
				\cite{jiangJointModelTask2022a},
				\cite{renEdgeAssistedDistributedDNN2020},
				\cite{zhangAutodidacticNeurosurgeonCollaborative2021a},
				\cite{DynamicSplitComputing2022},
			}
			]
			\addlegendimage{empty legend}
			\addlegendentry{Devices}
			\addplot[nodes near coords,fill=my_blue]
			table[x=x, y=y] {
				x y label
				1 23.93 \cite{liuLoADPartLoadAwareDynamic2022}
				2 23 \cite{xuDeepWearAdaptiveLocal2020}
				3 19.3 \cite{chenAcceleratingDNNInference2021a}
				4 6.78 \cite{zhouDynamicPathBased2021a}
				6 6.6 \cite{wangADDAAdaptiveDistributed2019}
				8 6.45 \cite{huDynamicAdaptiveDNN2019a}
				9 5.64 \cite{zhangRealTimeCooperativeDeep2020}
				11 2.97 \cite{liuEEAIEndedgeArchitecture2021a}
				12 2.5 \cite{jiangJointModelTask2022a}
				13 1.66 \cite{renEdgeAssistedDistributedDNN2020}
				14 1.6 \cite{zhangAutodidacticNeurosurgeonCollaborative2021a}
				15 1.29 \cite{DynamicSplitComputing2022}
			};
			\addlegendentry{2}
			\addplot[nodes near coords,fill=my_red]
			table[x=x, y=y] {
				x y label
				10 3.13 \cite{parasharProcessorPipeliningMethod2020a}
				5 6.67 \cite{heidariCAMDNNContentAwareMapping2022}
				7 6.58 \cite{yangCNNPCEndEdgeCloudCollaborative2022}
			};
			\addlegendentry{3}
		\end{axis}
	\end{tikzpicture}
	\Description{A bar plot showing the best relative latency speedup reported by each horizontal partitioning paper. The corresponding reference is provided for each bar in the plot. A different colour is used to group papers according to the amount of devices used.}
	\caption{Best latency speedup (when compared against the case with only one edge device) reported for papers using horizontal partitioning}
	\label{fig:horizontal_speedup}
\end{figure}
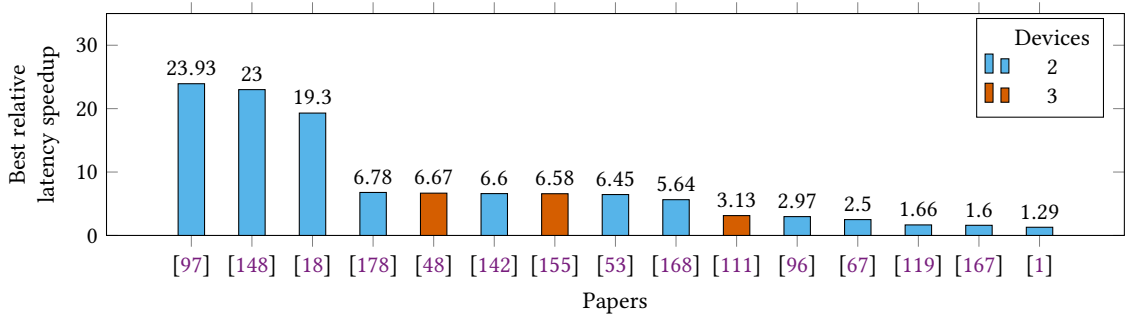

\begin{figure}[!ht]
	\centering
	\begin{tikzpicture}
		\begin{axis}[
			width=\textwidth,
			height=0.3\textwidth,
			xlabel=Papers,
			ylabel=Best relative \\ throughput speedup,
			ylabel style={align=center},
			ybar,
			ymin=0,
			ymax=50,
			bar shift=0pt,
			xtick={1,2,3,4,5,6,7},
			xticklabels={
				\cite{almeidaDynODynamicOnloading2022a},
				\cite{chenAcceleratingDNNInference2021a},
				\cite{huDynamicAdaptiveDNN2019a},
				\cite{leeSplittableDNNBasedObject2021a},
				\cite{huEnablePipelineProcessing2021a},
				\cite{koEdgeHostPartitioningDeep2018},
				\cite{parthasarathyDEFERDistributedEdge2022a},
			}
			]
			\addlegendimage{empty legend}
			\addlegendentry{Devices}
			\addplot[nodes near coords,fill=my_blue]
			table[x=x, y=y] {
				x y label
				1 35.5 \cite{almeidaDynODynamicOnloading2022a}
				2 16.5 \cite{chenAcceleratingDNNInference2021a}
				3 8.31 \cite{huDynamicAdaptiveDNN2019a}
				4 4.87 \cite{leeSplittableDNNBasedObject2021a}
				5 2.8 \cite{huEnablePipelineProcessing2021a}
				6 2.5 \cite{koEdgeHostPartitioningDeep2018}
			};
			\addlegendentry{2}
			\addplot[nodes near coords,fill=my_red]
			table[x=x, y=y] {
				x y label
				7 2 \cite{parthasarathyDEFERDistributedEdge2022a}
			};
			\addlegendentry{8}
		\end{axis}
	\end{tikzpicture}
	\Description{A bar plot showing the best relative throughput speedup reported by each horizontal partitioning paper. The corresponding reference is provided for each bar in the plot. A different colour is used to group papers according to the amount of devices used.}
	\caption{Best throughput speedup (when compared against the case with only one edge device) reported for papers using horizontal partitioning}
	\label{fig:horizontal_speedup_throughput}
\end{figure}
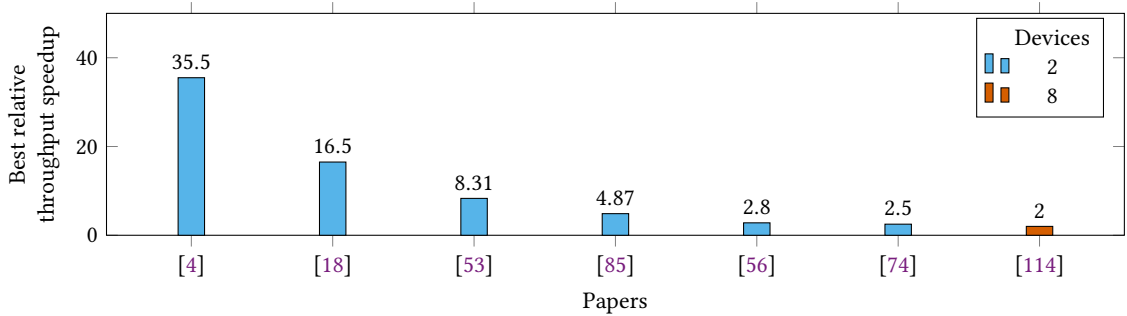

\subsubsection{Trends and challenges}

Figure \ref{fig:vertical_speedup} presents the best relative latency speedup reported for papers using a vertical partition plotted against the number of parallel devices used to achieve that speed. As expected, the paper that achieves the highest speedup is one that uses layer fusion: CoEdge \cite{zengCoEdgeCooperativeDNN2021a}. Interestingly, almost all other fusion papers rank \textit{below} those that do not use layer fusion. This can lead to two different conclusions. One of them is that layer fusion, although promising, does not seem to be the definitive factor that automatically improves the solutions obtained by distribution algorithms. However, because these papers always rank better in their own papers compared to non-fusion papers under the same conditions, this is highly unlikely. The second (and more likely) explanation is that the test setup and parameters of the distributed system significantly influence the decisions found by the distribution algorithm (this is an important topic that will be discussed in Section \ref{section:nn_topologies}). 

Figure \ref{fig:energy_savings} presents the best reported energy savings, calculated as the energy of the inference execution on one device divided by the energy of the execution on a distributed manner. As it can be appreciated from the number of datapoints, this is not usually reported. Future papers should report this metric in order to be able to easily compare papers optimizing for energy between them. Another promising research direction is the optimization of the consumed energy for papers using vertical partitioning, which is still a gap in the literature.

Finally, figures \ref{fig:horizontal_speedup} and \ref{fig:horizontal_speedup_throughput} present the best relative latency and throughput speedup for papers using horizontal partitioning. 

Table \ref{tab:target_to_optimize} shows a clear focus on optimizing the time-related metrics of both latency and throughput. Given current $CO_2$ emissions concerns, we expect to see a natural change in the research landscape of distributed inference towards optimizing energy. Most studies also focus on optimizing only one type of metric, but generating a multi-objective optimization problem presents itself as a promising research direction. We also expect to see a greater focus on optimizing for privacy given current data protection laws and concerns. The final promising research path we detected is the development of new evaluation metrics to make the quality of these papers independent from their test environments.



\subsection{Devices cost model}
\label{section:devices_cost_model}

In distributed inference papers, a device cost model is almost always needed if the partitioning and allocation need to be selected automatically. In that case, it is necessary to have a model that can predict the cost (one of the metrics described in section \ref{section:target_to_optimize}) of executing a layer with specific parameters on a particular device. We detected that distributed inference papers can also be grouped depending on which technique they use to model their devices.

A number of papers use analytical models to represent each available device. These models can be as simple as the ones used by CoopAI \cite{yangCooperativeDistributedDeep2021a} which uses the amount of GOP of a particular layer divided by the GOP/s of each device to find the cost, or as complex as the ones used by Super-LIP \cite{jiangAchievingSuperLinearSpeedup2019}, which use an analytical to describe the entire hardware architecture running on each FPGA. 

Other papers use simulators like Scale-SIM \cite{DBLP:journals/corr/abs-1811-02883} or Timeloop \cite{8695666} to model with more or less degree of detail the architecture of their devices. Depending on the simulator, obtaining the cost of each layer and each partition can be costly in terms of time, but remains an interesting option given their accuracy.

Offline profiling is one of the most accurate methods to obtain the cost of executing a particular layer on each available device. We refer to this method as \textit{offline} because the measurement of each layer is executed only once, and not in a continuous manner under the actual operating conditions. But measuring each layer and each partition on all available devices still remains a costly operation. This is why this technique is normally used in conjunction with regression models: a subset of layers with different parameters are first profiled, a regression model is trained using these measurements as inputs, and the regression model is used to predict the behaviour of the rest of the needed layers and partitions.

On the other hand, the online profiling option is of particular interest for papers implementing a dynamic distributed system, as they not only adapt to the communication channels state but also to computing capabilities changes (for example, because of other workloads running on the same device). This method is used when a periodic update of the costs is needed by the algorithm. At periodic intervals, the actual cost on each device is measured (or traced) and this is used to guide the selection of new distribution options.

A significant amount of papers do not provide information or do not need a device model, normally because they use heuristics or manual decisions on how to partition the DNN, and do not rely on an automated algorithm to find the best possible solution.

{\footnotesize
\begin{table}[htbp]
	\caption{Papers categorized according to the kind of cost model they used to describe their devices}
	\begin{center}
		\begin{tabularx}{\textwidth}{>{\hsize=.3\hsize}X>{\hsize=.7\hsize}X} \toprule
			Category & Papers \\
			\midrule
			Analytical & \cite{chenAcceleratingDNNInference2021a,jiangAchievingSuperLinearSpeedup2019,niuAdaptiveDeviceEdgeCoInference2022,zhangCommunicationComputationEfficientDeviceEdge2021,yangCooperativeDistributedDeep2021a,khanDistributedInferenceResourceConstrained2022,yangEdgeComputingNetworking2021a,xueEdgeLDLocallyDistributed2020a,huangEnablingLowLatency2021a,jinEnergyAwareWorkloadAllocation2019,kroukaEnergyEfficientModelCompression2021,pachecoInferenceTimeOptimization2020a,yangIntelliEyeUAVTracking2019,heJointDNNPartition2020a,dongJointOptimizationDNN2021a,leeNeuralArchitectureSearch2021a,yangOffloadingOptimizationEdge2021a,deyPartitioningCNNModels2018a,luResourceEfficientDistributedDeep2022,baccourRLPDNNReinforcementLearning2021a,fuSplitComputingVideo2022a,yangEfficientInferenceAdaptively2021a,baccourDistPrivacyPrivacyAwareDistributed2020,hsuCooperativeConvolutionalNeural2020,renEdgeAssistedDistributedDNN2020,changUltraLowLatencyDistributedDeep2019} \\
			Simulator & \cite{banitalebi-dehkordiAutoSplitGeneralFramework2021,kressHardwareawarePartitioningConvolutional2022a} \\
			Offline & \cite{miaoAdaptiveDNNPartition2020a,dagliAxoNNEnergyAwareExecution2022,yangCNNPCEndEdgeCloudCollaborative2022,zengCoEdgeCooperativeDNN2021a,yunCooperativeInferenceDNNs2022a,houDistrEdgeSpeedingConvolutional2022a,gacoinDistributingDeepNeural2019a,zhouDynamicPathBased2021a,wangDynamicResourceAllocation2021a,DynamicSplitComputing2022,liEnablingRealtimeAI2022a,huFastAccurateStreaming2020a,songIndustrialVisionOptimization2021a,liJALADJointAccuracyAnd2018,jiangJointModelTask2022a,choiLegionTailoringGrouped2021a,haoMultiAgentCollaborativeInference2022,jiNovelAdaptiveDNN2022a,deyOffloadedExecutionDeep2019a,parasharProcessorPipeliningMethod2020a,liReceptiveFieldbasedSegmentation2022a,SplitComputingDNN2022,zhangRealTimeCooperativeDeep2020,hadidiDistributedPerceptionCollaborative2018a,zhouAAIoTAcceleratingArtificial2019,eshratifarJointDNNEfficientTraining2021,huangCLIOEnablingAutomatic2020,eshratifarBottleNetDeepLearning2019} \\
			Offline + regression & \cite{zhouAdaptiveParallelExecution2019a,zengBoomerangOnDemandCooperative2019,jiaCoDLEfficientCPUGPU2022a,vanishreeCoInAcceleratedCNN2020a,huDistributedInferenceDeep2022a,zhangDynamicDNNDecomposition2021a,almeidaDynODynamicOnloading2022a,liEdgeAIOnDemand2020,liuEEAIEndedgeArchitecture2021a,huEnablePipelineProcessing2021a,xuEOPEfficientOperator2022a,wuHiTDLHighThroughputDeep2022,fangJointArchitectureDesign2022a,fuJointOptimizationData2021a,xuDeepWearAdaptiveLocal2020,maoMoDNNLocalDistributed2017a,wangADDAAdaptiveDistributed2019,kangNeurosurgeonCollaborativeIntelligence2017a,laskaridisSPINNSynergisticProgressive2020,jeongIONNIncrementalOffloading2018} \\
			Online & \cite{zhangAdaptiveDistributedConvolutional2020a,samikwaAdaptiveEarlyExit2022a,zhangAutodidacticNeurosurgeonCollaborative2021a,kimAutoScaleEnergyEfficiency2020a,heidariCAMDNNContentAwareMapping2022,luoCloudEdgeCollaborativeIntelligent2021,xueDDPQNEfficientDNN2022a,huangDeeParHybridDeviceEdgeCloud2019,zhangDeepSlicingCollaborativeAdaptive2021a,mohammedDistributedInferenceAcceleration2020a,huDynamicAdaptiveDNN2019a,zhangElfAccelerateHighResolution2021,liuLoADPartLoadAwareDynamic2022,shiPrivacyAwareEdgeComputing2019,tuliSplitPlaceAIAugmented2022a,hadidiCollaborativeInferencingDeep2020a,laskaridisSPINNSynergisticProgressive2020} \\
			Not needed / no information & \cite{wangFastEnergySavingNeural2021,leeSplittableDNNBasedObject2021a,zhangAccelerateDeepLearning2020a,changEfficientDistributedDeep2019a,zouCAPCommunicationAwareAutomated2022,baeCapellaCustomizingPerception2019a,shaoCommunicationComputationTradeoffResourceConstrained2020,sbaiCutDistilEncode2021a,zhaoDeepThingsDistributedAdaptive2018a,parthasarathyDEFERDistributedEdge2022a,sahuDENNIDistributedNeural2021a,liDistributedDeepLearning2022a,naveenLowLatencyDeep2021a,jeongOptimalPartitioningDistributed2021a,fangTeamNetCollaborativeInference2019a,camposdeoliveiraPartitioningConvolutionalNeural2018,teerapittayanonDistributedDeepNeural2017,koEdgeHostPartitioningDeep2018,matsubaraDistilledSplitDeep2019,shaoBottleNetEndtoEndApproach2020} \\
			\bottomrule
		\end{tabularx}
		\label{tab:devices_cost_model}
	\end{center}
\end{table}
}

\subsubsection{Trends and challenges}

Table \ref{tab:devices_cost_model} presents the reviewed papers categorized according to the kind of cost model used for their device. There is a clear research gap in the usage of simulators to model the device's cost that should be addressed in the following papers. It can also be seen that offline profiling methods, also when using regression models, are the most commonly used methods.

A promising research path is the comparison of different device cost models, and the evaluation of how these affect the distribution strategies found. This could give more information about how accurate the device model actually needs to be.´

\subsection{Other categorizations}

\subsubsection{DNN topologies and dataset}
\label{section:nn_topologies}

It is important to take into account the topology of the DNN that is being distributed because the number of layers and the size of the intermediate feature maps greatly impact the partition decisions. But the dataset used to train the DNN is also important, because the input size also modifies the size of the intermediate feature maps. These two aspects also modify the amount of GOP needed for an inference pass, which in turn modifies the algorithm's decisions. As such, when comparing these kinds of papers, one should be aware of both aspects in order to be able to do a fair comparison. 

This is why it's so important to use publicly available DNN topologies, which improve the reproducibility of these kinds of papers. As such, custom DNN topologies can be perfect to solve particular use cases, but from a research point of view, they present low reproducibility. As it can be seen in Figure \ref{fig:topology_count}, VGG and its variants are the most commonly used networks. One reason for this is its simple, sequential-like structure, which does not require complex partitioning decisions. The same could be said about AlexNet. But interestingly, there is quite a lot of research focused on distributing more complex architectures like ResNet, MobileNet, YOLO or Inception. These architectures present challenges because of their branches and residual connections, which require specific techniques to distribute them. Table \ref{tab:nn_used} provides an overview of the papers that use each DNN architecture.

\begin{figure}[htbp]
	\centering
	\begin{tikzpicture}
		\begin{axis}[
			width=\textwidth,
			height=0.3\textwidth,
			xlabel=Topology,
			ylabel=Paper count,
			ybar,
			ymin=0,
			xtick=data,
			symbolic x coords={
				VGG,
				ResNet,
				AlexNet,
				MobileNet,
				Yolo,
				Inception
			},
			],
			\addplot[fill=my_blue] table[x=index,y=Count,col sep=comma] {./data/topology_count.csv};
		\end{axis}
	\end{tikzpicture}
	\Description{A bar plot showing how many of the surveyed papers use each neural network architecture on their evaluation sections. VGG, ResNet and AlexNet are the most commonly used.}
	\caption{Amount of papers using versions of the most common DNN topologies}
	\label{fig:topology_count}
\end{figure}
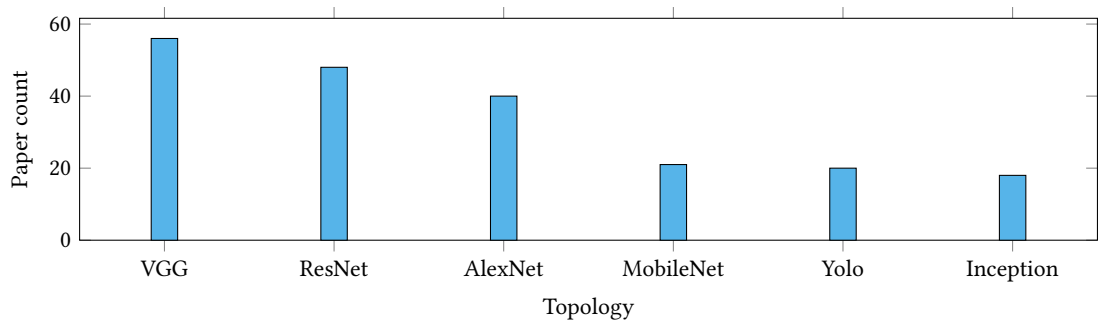

As was mentioned earlier, the dataset used to train the DNN also plays an important role in the decisions taken by the distribution algorithm. Of the 112 surveyed papers, 48 don't talk about the selected dataset used to train the DNN distributed in the experiments section, but \cite{zhangAutodidacticNeurosurgeonCollaborative2021a,zhangDeepSlicingCollaborativeAdaptive2021a,liuLoADPartLoadAwareDynamic2022,parasharProcessorPipeliningMethod2020a,liReceptiveFieldbasedSegmentation2022a} provide the size of the input image, which should be enough information to be able to replicate the test conditions. As shown in Table \ref{tab:dataset_used}, there is a clear trend to use models pre-trained on ImageNet \cite{NIPS2012_c399862d}. 

{\footnotesize
\begin{table}[htbp]
	\caption{Papers categorized according to the dataset used to train the distributed networks}
	\begin{center}
		\begin{tabularx}{\textwidth}{>{\hsize=.3\hsize}X>{\hsize=.7\hsize}X} \toprule
			Dataset & Papers \\
			\midrule
			ImageNet \cite{imagenet_cvpr09} & \cite{chenAcceleratingDNNInference2021a, zhangAdaptiveDistributedConvolutional2020a, banitalebi-dehkordiAutoSplitGeneralFramework2021, zouCAPCommunicationAwareAutomated2022, baeCapellaCustomizingPerception2019a, yangCNNPCEndEdgeCloudCollaborative2022, zengCoEdgeCooperativeDNN2021a, vanishreeCoInAcceleratedCNN2020a, yangCooperativeDistributedDeep2021a, sbaiCutDistilEncode2021a, parthasarathyDEFERDistributedEdge2022a, sahuDENNIDistributedNeural2021a, khanDistributedInferenceResourceConstrained2022, huDistributedInferenceDeep2022a, zhangDynamicDNNDecomposition2021a, zhouDynamicPathBased2021a, almeidaDynODynamicOnloading2022a, xueEdgeLDLocallyDistributed2020a, pachecoInferenceTimeOptimization2020a, liJALADJointAccuracyAnd2018, yangOffloadingOptimizationEdge2021a, deyPartitioningCNNModels2018a, zhouAAIoTAcceleratingArtificial2019, maoMoDNNLocalDistributed2017a, laskaridisSPINNSynergisticProgressive2020, huangCLIOEnablingAutomatic2020, eshratifarBottleNetDeepLearning2019} \\
			CIFAR-10 \cite{krizhevsky2009learning} & \cite{niuAdaptiveDeviceEdgeCoInference2022, changEfficientDistributedDeep2019a, zengBoomerangOnDemandCooperative2019, zouCAPCommunicationAwareAutomated2022, luoCloudEdgeCollaborativeIntelligent2021, vanishreeCoInAcceleratedCNN2020a, zhangCommunicationComputationEfficientDeviceEdge2021, shaoCommunicationComputationTradeoffResourceConstrained2020, yunCooperativeInferenceDNNs2022a, gacoinDistributingDeepNeural2019a, zhouDynamicPathBased2021a, wangDynamicResourceAllocation2021a, liEdgeAIOnDemand2020, huEnablePipelineProcessing2021a, huangEnablingLowLatency2021a, huFastAccurateStreaming2020a, jiangJointModelTask2022a, leeNeuralArchitectureSearch2021a, jeongOptimalPartitioningDistributed2021a, baccourRLPDNNReinforcementLearning2021a, tuliSplitPlaceAIAugmented2022a, fangTeamNetCollaborativeInference2019a, renEdgeAssistedDistributedDNN2020, wangADDAAdaptiveDistributed2019, laskaridisSPINNSynergisticProgressive2020, huangCLIOEnablingAutomatic2020, shaoBottleNetEndtoEndApproach2020} \\
			MNIST \cite{lecun-mnisthandwrittendigit-2010} & \cite{zouCAPCommunicationAwareAutomated2022,vanishreeCoInAcceleratedCNN2020a, huangDeeParHybridDeviceEdgeCloud2019, kroukaEnergyEfficientModelCompression2021, jiangJointModelTask2022a, baccourRLPDNNReinforcementLearning2021a, tuliSplitPlaceAIAugmented2022a, fangTeamNetCollaborativeInference2019a, baccourDistPrivacyPrivacyAwareDistributed2020} \\
			CIFAR-100 \cite{krizhevsky2009learning} & \cite{huFastAccurateStreaming2020a, jeongOptimalPartitioningDistributed2021a, tuliSplitPlaceAIAugmented2022a, renEdgeAssistedDistributedDNN2020, laskaridisSPINNSynergisticProgressive2020, shaoBottleNetEndtoEndApproach2020} \\
			COCO \cite{lin2014microsoft} & \cite{leeSplittableDNNBasedObject2021a, banitalebi-dehkordiAutoSplitGeneralFramework2021, jiNovelAdaptiveDNN2022a, fuSplitComputingVideo2022a} \\
			BDD100k \cite{DBLP:journals/corr/abs-1805-04687} & \cite{mohammedDistributedInferenceAcceleration2020a, huDynamicAdaptiveDNN2019a, zhangRealTimeCooperativeDeep2020} \\
			PASCAL VOC \cite{Everingham10} & \cite{zhangAdaptiveDistributedConvolutional2020a, yangCNNPCEndEdgeCloudCollaborative2022} \\
			NEU-CLS \cite{8709818} & \cite{fangJointArchitectureDesign2022a} \\
			KITTI \cite{Menze2015CVPR} & \cite{zhangElfAccelerateHighResolution2021} \\
			Intel Image Classification \cite{Bansal_2019} & \cite{luResourceEfficientDistributedDeep2022} \\
			Stanford CARs \cite{KrauseStarkDengFei-Fei_3DRR2013} & \cite{baccourRLPDNNReinforcementLearning2021a} \\
			FashionMNIST \cite{xiao2017/online} & \cite{tuliSplitPlaceAIAugmented2022a} \\
			CELEBA \cite{liu2015faceattributes} & \cite{baccourDistPrivacyPrivacyAwareDistributed2020} \\
			PoseTrack \cite{8578640} & \cite{zhangElfAccelerateHighResolution2021} \\
			MOTS \cite{8953401} & \cite{zhangElfAccelerateHighResolution2021} \\
			PCB \cite{DBLP:journals/corr/abs-1901-08204} & \cite{yangEdgeComputingNetworking2021a} \\
			CamVid \cite{BrostowFC:PRL2008} & \cite{zhangAdaptiveDistributedConvolutional2020a} \\
			Caltech 101 \cite{li_andreeto_ranzato_perona_2022} & \cite{zhangAdaptiveDistributedConvolutional2020a} \\
			PETS09 \cite{5399556} & \cite{teerapittayanonDistributedDeepNeural2017} \\
			UCI \cite{9060182} & \cite{sahuDENNIDistributedNeural2021a} \\
			\bottomrule
		\end{tabularx}
		\label{tab:dataset_used}
	\end{center}
\end{table}
}

{\footnotesize
\begin{table}[htbp]
	\caption{Papers categorized according to the evaluated DNN topologies used for experimental results and comparisons}
	\begin{center}
		\begin{tabularx}{\textwidth}{>{\hsize=.2\hsize}X>{\hsize=.8\hsize}X} \toprule
			Network & Papers \\
			\midrule
			VGG \cite{simonyan2015a} & \cite{chenAcceleratingDNNInference2021a, jiangAchievingSuperLinearSpeedup2019, zhangAdaptiveDistributedConvolutional2020a, samikwaAdaptiveEarlyExit2022a, zhouAdaptiveParallelExecution2019a, zhangAutodidacticNeurosurgeonCollaborative2021a, dagliAxoNNEnergyAwareExecution2022, zouCAPCommunicationAwareAutomated2022, baeCapellaCustomizingPerception2019a, yangCNNPCEndEdgeCloudCollaborative2022, jiaCoDLEfficientCPUGPU2022a, zengCoEdgeCooperativeDNN2021a, yangCooperativeDistributedDeep2021a, sbaiCutDistilEncode2021a, xueDDPQNEfficientDNN2022a, huangDeeParHybridDeviceEdgeCloud2019, zhangDeepSlicingCollaborativeAdaptive2021a, parthasarathyDEFERDistributedEdge2022a, houDistrEdgeSpeedingConvolutional2022a, liDistributedDeepLearning2022a, mohammedDistributedInferenceAcceleration2020a, khanDistributedInferenceResourceConstrained2022, huDistributedInferenceDeep2022a, huDynamicAdaptiveDNN2019a, zhangDynamicDNNDecomposition2021a, wangDynamicResourceAllocation2021a, almeidaDynODynamicOnloading2022a, xueEdgeLDLocallyDistributed2020a, huangEnablingLowLatency2021a, jinEnergyAwareWorkloadAllocation2019, songIndustrialVisionOptimization2021a, liJALADJointAccuracyAnd2018, fangJointArchitectureDesign2022a, fuJointOptimizationData2021a, choiLegionTailoringGrouped2021a, liuLoADPartLoadAwareDynamic2022, haoMultiAgentCollaborativeInference2022, jiNovelAdaptiveDNN2022a, deyPartitioningCNNModels2018a, parasharProcessorPipeliningMethod2020a, liReceptiveFieldbasedSegmentation2022a, baccourRLPDNNReinforcementLearning2021a, yangEfficientInferenceAdaptively2021a, baccourDistPrivacyPrivacyAwareDistributed2020, hadidiDistributedPerceptionCollaborative2018a, hadidiCollaborativeInferencingDeep2020a, maoMoDNNLocalDistributed2017a, renEdgeAssistedDistributedDNN2020, eshratifarJointDNNEfficientTraining2021, koEdgeHostPartitioningDeep2018, wangADDAAdaptiveDistributed2019, kangNeurosurgeonCollaborativeIntelligence2017a, laskaridisSPINNSynergisticProgressive2020, huangCLIOEnablingAutomatic2020, shaoBottleNetEndtoEndApproach2020, eshratifarBottleNetDeepLearning2019} \\
			ResNet \cite{7780459} & \cite{chenAcceleratingDNNInference2021a, zhangAdaptiveDistributedConvolutional2020a, zhouAdaptiveParallelExecution2019a, changEfficientDistributedDeep2019a, banitalebi-dehkordiAutoSplitGeneralFramework2021, zhangAutodidacticNeurosurgeonCollaborative2021a, kimAutoScaleEnergyEfficiency2020a, dagliAxoNNEnergyAwareExecution2022, zouCAPCommunicationAwareAutomated2022, baeCapellaCustomizingPerception2019a, luoCloudEdgeCollaborativeIntelligent2021, yangCNNPCEndEdgeCloudCollaborative2022, vanishreeCoInAcceleratedCNN2020a, zhangCommunicationComputationEfficientDeviceEdge2021, shaoCommunicationComputationTradeoffResourceConstrained2020, yunCooperativeInferenceDNNs2022a, xueDDPQNEfficientDNN2022a, zhangDeepSlicingCollaborativeAdaptive2021a, parthasarathyDEFERDistributedEdge2022a, houDistrEdgeSpeedingConvolutional2022a, mohammedDistributedInferenceAcceleration2020a, gacoinDistributingDeepNeural2019a, huDynamicAdaptiveDNN2019a, zhangDynamicDNNDecomposition2021a, zhouDynamicPathBased2021a, almeidaDynODynamicOnloading2022a, huangEnablingLowLatency2021a, liEnablingRealtimeAI2022a, kressHardwareawarePartitioningConvolutional2022a, wuHiTDLHighThroughputDeep2022, liJALADJointAccuracyAnd2018, fangJointArchitectureDesign2022a, liuLoADPartLoadAwareDynamic2022, haoMultiAgentCollaborativeInference2022, parasharProcessorPipeliningMethod2020a, luResourceEfficientDistributedDeep2022, tuliSplitPlaceAIAugmented2022a, zhangRealTimeCooperativeDeep2020, hsuCooperativeConvolutionalNeural2020, hadidiCollaborativeInferencingDeep2020a, renEdgeAssistedDistributedDNN2020, eshratifarJointDNNEfficientTraining2021, koEdgeHostPartitioningDeep2018, laskaridisSPINNSynergisticProgressive2020, jeongIONNIncrementalOffloading2018, huangCLIOEnablingAutomatic2020, shaoBottleNetEndtoEndApproach2020, eshratifarBottleNetDeepLearning2019} \\
			AlexNet \cite{NIPS2012_c399862d} & \cite{zhangAccelerateDeepLearning2020a, chenAcceleratingDNNInference2021a, jiangAchievingSuperLinearSpeedup2019, samikwaAdaptiveEarlyExit2022a, niuAdaptiveDeviceEdgeCoInference2022, dagliAxoNNEnergyAwareExecution2022, zengBoomerangOnDemandCooperative2019, zouCAPCommunicationAwareAutomated2022, baeCapellaCustomizingPerception2019a, zengCoEdgeCooperativeDNN2021a, yangCooperativeDistributedDeep2021a, xueDDPQNEfficientDNN2022a, zhangDeepSlicingCollaborativeAdaptive2021a, mohammedDistributedInferenceAcceleration2020a, huDynamicAdaptiveDNN2019a, zhangDynamicDNNDecomposition2021a, wangDynamicResourceAllocation2021a, liEdgeAIOnDemand2020, liuEEAIEndedgeArchitecture2021a, huEnablePipelineProcessing2021a, jinEnergyAwareWorkloadAllocation2019, pachecoInferenceTimeOptimization2020a, jiangJointModelTask2022a, fuJointOptimizationData2021a, dongJointOptimizationDNN2021a, liuLoADPartLoadAwareDynamic2022, jiNovelAdaptiveDNN2022a, deyOffloadedExecutionDeep2019a, shiPrivacyAwareEdgeComputing2019, zhangRealTimeCooperativeDeep2020, hadidiDistributedPerceptionCollaborative2018a, hadidiCollaborativeInferencingDeep2020a, zhouAAIoTAcceleratingArtificial2019, renEdgeAssistedDistributedDNN2020, changUltraLowLatencyDistributedDeep2019, eshratifarJointDNNEfficientTraining2021, koEdgeHostPartitioningDeep2018, wangADDAAdaptiveDistributed2019, kangNeurosurgeonCollaborativeIntelligence2017a, jeongIONNIncrementalOffloading2018} \\
			MobileNet \cite{DBLP:journals/corr/abs-1905-02244} & \cite{banitalebi-dehkordiAutoSplitGeneralFramework2021, kimAutoScaleEnergyEfficiency2020a, heidariCAMDNNContentAwareMapping2022, zouCAPCommunicationAwareAutomated2022, yangCNNPCEndEdgeCloudCollaborative2022, zengCoEdgeCooperativeDNN2021a, vanishreeCoInAcceleratedCNN2020a, sbaiCutDistilEncode2021a, khanDistributedInferenceResourceConstrained2022, almeidaDynODynamicOnloading2022a, xuEOPEfficientOperator2022a, huFastAccurateStreaming2020a, wuHiTDLHighThroughputDeep2022, haoMultiAgentCollaborativeInference2022, SplitComputingDNN2022, tuliSplitPlaceAIAugmented2022a, xuDeepWearAdaptiveLocal2020, renEdgeAssistedDistributedDNN2020, laskaridisSPINNSynergisticProgressive2020, jeongIONNIncrementalOffloading2018} \\
			Yolo \cite{DBLP:journals/corr/RedmonDGF15} & \cite{leeSplittableDNNBasedObject2021a, jiangAchievingSuperLinearSpeedup2019, zhangAdaptiveDistributedConvolutional2020a, zhouAdaptiveParallelExecution2019a, banitalebi-dehkordiAutoSplitGeneralFramework2021, zhangAutodidacticNeurosurgeonCollaborative2021a, jiaCoDLEfficientCPUGPU2022a, yangCooperativeDistributedDeep2021a, zhaoDeepThingsDistributedAdaptive2018a, houDistrEdgeSpeedingConvolutional2022a, huDistributedInferenceDeep2022a, huDynamicAdaptiveDNN2019a, yangEdgeComputingNetworking2021a, xuEOPEfficientOperator2022a, choiLegionTailoringGrouped2021a, naveenLowLatencyDeep2021a, fuSplitComputingVideo2022a, yangEfficientInferenceAdaptively2021a, zhangRealTimeCooperativeDeep2020, changUltraLowLatencyDistributedDeep2019,zhangDynamicDNNDecomposition2021a, zhangRealTimeCooperativeDeep2020} \\
			Inception \cite{DBLP:journals/corr/SzegedyVISW15} & \cite{miaoAdaptiveDNNPartition2020a, kimAutoScaleEnergyEfficiency2020a, heidariCAMDNNContentAwareMapping2022, houDistrEdgeSpeedingConvolutional2022a, zhangDynamicDNNDecomposition2021a, zhouDynamicPathBased2021a, wangDynamicResourceAllocation2021a, almeidaDynODynamicOnloading2022a, huangEnablingLowLatency2021a, liEnablingRealtimeAI2022a, xuEOPEfficientOperator2022a, wuHiTDLHighThroughputDeep2022, deyOffloadedExecutionDeep2019a, deyPartitioningCNNModels2018a, parasharProcessorPipeliningMethod2020a, tuliSplitPlaceAIAugmented2022a, laskaridisSPINNSynergisticProgressive2020, jeongIONNIncrementalOffloading2018} \\
			GoogLeNet \cite{DBLP:journals/corr/SzegedyLJSRAEVR14} & \cite{banitalebi-dehkordiAutoSplitGeneralFramework2021, dagliAxoNNEnergyAwareExecution2022, zengCoEdgeCooperativeDNN2021a, vanishreeCoInAcceleratedCNN2020a, xueDDPQNEfficientDNN2022a, zhangDeepSlicingCollaborativeAdaptive2021a, kressHardwareawarePartitioningConvolutional2022a, zhangRealTimeCooperativeDeep2020, xuDeepWearAdaptiveLocal2020, jeongIONNIncrementalOffloading2018} \\
			SqueezeNet \cite{DBLP:journals/corr/IandolaMAHDK16} & \cite{jiangAchievingSuperLinearSpeedup2019, heidariCAMDNNContentAwareMapping2022, vanishreeCoInAcceleratedCNN2020a, huangEnablingLowLatency2021a, kressHardwareawarePartitioningConvolutional2022a, liuLoADPartLoadAwareDynamic2022, deyOffloadedExecutionDeep2019a} \\
			LeNet \cite{726791} & \cite{huangDeeParHybridDeviceEdgeCloud2019, jiangJointModelTask2022a, jiNovelAdaptiveDNN2022a, baccourRLPDNNReinforcementLearning2021a, baccourDistPrivacyPrivacyAwareDistributed2020, camposdeoliveiraPartitioningConvolutionalNeural2018, huangCLIOEnablingAutomatic2020} \\
			Custom DNN & \cite{kroukaEnergyEfficientModelCompression2021, yangIntelliEyeUAVTracking2019, yangOffloadingOptimizationEdge2021a, baccourRLPDNNReinforcementLearning2021a, fangTeamNetCollaborativeInference2019a} \\
			PoseNet \cite{DBLP:journals/corr/KendallGC15} & \cite{jiaCoDLEfficientCPUGPU2022a, DynamicSplitComputing2022, liEnablingRealtimeAI2022a, SplitComputingDNN2022} \\
			NiN \cite{lin2013network} & \cite{chenAcceleratingDNNInference2021a, mohammedDistributedInferenceAcceleration2020a, huDynamicAdaptiveDNN2019a, eshratifarJointDNNEfficientTraining2021} \\
			DenseNet \cite{8099726} & \cite{liEnablingRealtimeAI2022a, parasharProcessorPipeliningMethod2020a, matsubaraDistilledSplitDeep2019} \\
			HAR \cite{9835797} & \cite{ sahuDENNIDistributedNeural2021a, xuDeepWearAdaptiveLocal2020} \\
			Xception \cite{DBLP:journals/corr/Chollet16a} & \cite{liuLoADPartLoadAwareDynamic2022, hadidiCollaborativeInferencingDeep2020a} \\
			DeepFace \cite{BMVC2015_41} & \cite{huangDeeParHybridDeviceEdgeCloud2019, kangNeurosurgeonCollaborativeIntelligence2017a} \\
			BNN \cite{DBLP:journals/corr/CourbariauxBD15} & \cite{teerapittayanonDistributedDeepNeural2017} \\
			Conv-TasNet \cite{8707065} & \cite{wuHiTDLHighThroughputDeep2022} \\
			EfficientNet \cite{DBLP:journals/corr/abs-1905-11946} & \cite{wuHiTDLHighThroughputDeep2022} \\
			MBNN \cite{9170543}& \cite{wangFastEnergySavingNeural2021} \\
			WRN \cite{DBLP:journals/corr/ZagoruykoK16} & \cite{jeongOptimalPartitioningDistributed2021a} \\
			C3D \cite{7410867} & \cite{hadidiCollaborativeInferencingDeep2020a} \\
			DeepSense \cite{DBLP:journals/corr/YaoHZZA16} & \cite{xuDeepWearAdaptiveLocal2020} \\
			Deeplab \cite{7913730} & \cite{liEnablingRealtimeAI2022a} \\
			WaveNet \cite{DBLP:journals/corr/OordDZSVGKSK16} & \cite{xuDeepWearAdaptiveLocal2020} \\
			OverFeat \cite{Sermanet2013OverFeatIR} & \cite{eshratifarJointDNNEfficientTraining2021} \\
			Deep Speech \cite{DBLP:journals/corr/HannunCCCDEPSSCN14} & \cite{eshratifarJointDNNEfficientTraining2021} \\
			Kaldi \cite{ravanelli2019pytorchkaldi} & \cite{kangNeurosurgeonCollaborativeIntelligence2017a} \\
			DeepEar \cite{10.1145/2750858.2804262} & \cite{xuDeepWearAdaptiveLocal2020} \\
			FoveaBox \cite{9123553} & \cite{zhangElfAccelerateHighResolution2021} \\
			FaceNet \cite{7298682} & \cite{liEnablingRealtimeAI2022a} \\
			RetinaNet \cite{8237586} & \cite{zhangElfAccelerateHighResolution2021} \\
			FCN \cite{7298965} & \cite{zhangAdaptiveDistributedConvolutional2020a} \\
			CharCNN \cite{NIPS2015_250cf8b5} & \cite{zhangAdaptiveDistributedConvolutional2020a} \\
			OpenPose \cite{DBLP:journals/corr/abs-1812-08008} & \cite{houDistrEdgeSpeedingConvolutional2022a} \\
			VoxelNet \cite{8578570} & \cite{houDistrEdgeSpeedingConvolutional2022a} \\
			CascadeRCNN \cite{8578742} & \cite{zhangElfAccelerateHighResolution2021} \\
			DynamicRCNN \cite{DBLP:journals/corr/abs-2004-06002} & \cite{zhangElfAccelerateHighResolution2021} \\
			FasterRCNN \cite{7485869} & \cite{zhangElfAccelerateHighResolution2021} \\
			FCOS \cite{9010746} & \cite{zhangElfAccelerateHighResolution2021} \\
			FreeAnchor \cite{9321141} & \cite{zhangElfAccelerateHighResolution2021} \\
			FSAF \cite{8953532} & \cite{zhangElfAccelerateHighResolution2021} \\
			MaskRCNN \cite{matterport_maskrcnn_2017} & \cite{zhangElfAccelerateHighResolution2021} \\
			NasFPN \cite{8954436} & \cite{zhangElfAccelerateHighResolution2021} \\
			SENNA \cite{DBLP:journals/corr/abs-1103-0398} & \cite{kangNeurosurgeonCollaborativeIntelligence2017a} \\
			\bottomrule
		\end{tabularx}
		\label{tab:nn_used}
	\end{center}
\end{table}
}

\subsubsection{Resulting DNN topology}
\label{section:resulting_nn_topology}

Although most papers take a pre-trained network and distribute it without changing its inherent structure, this survey identified a number of papers that modify the network's architecture and re-train it in order to make it more suitable for edge devices. We identify two kinds of papers: those which only transform an existing architecture, without changing its structure significantly (for example, by inserting compression layers or replacing particular layers), and those which generate a completely new DNN. Table \ref{tab:resulting_nn_topology} provides an overview of this categorization, together with the method used to transform the original DNN.

\cite{leeSplittableDNNBasedObject2021a} uses a feature reconstructor network in the cloud in order to avoid sending all features generated by the network running on the edge device. \cite{zhangAdaptiveDistributedConvolutional2020a} introduces the concept of \textit{Fully Decomposable Spatial Partition} (FDSP), which eliminates the inter-tile communication and synchronization problems presented in section \ref{section:vertical_partition}. This method generates independent tiles for the first convolutional layers of the network by partitioning the \ifm in a grid manner. Each tile is then padded with zeros on all its borders, which modifies the actual mathematical operation of the original, non-partitioned, convolutional layer, and can introduce accuracy losses. \cite{yangCNNPCEndEdgeCloudCollaborative2022} modifies the network with two methods. First, they use \textit{Identical Channel Pruning} (ICP) to reduce the amount of communicated data. Then, compression layers are added using \textit{Compression Rate Determination} (CRD). Several papers \cite{yunCooperativeInferenceDNNs2022a,sbaiCutDistilEncode2021a,jeongOptimalPartitioningDistributed2021a,matsubaraDistilledSplitDeep2019} use KD to generate a smaller network to be executed on the edge based on the knowledge extracted from a bigger more computationally or memory expensive teacher network. Finally, some papers like \cite{eshratifarBottleNetDeepLearning2019} propose to use Auto-encoders (AE) to reduce the size of the transmitted data between devices. 

{\footnotesize
\begin{table}[htbp]
	\caption{Papers categorized according to the resulting DNN topology}
	\begin{center}
		\begin{tabularx}{\textwidth}{>{\hsize=.25\hsize}X>{\hsize=.35\hsize}X>{\hsize=.4\hsize}X} \toprule
			Paper & Resulting DNN topology & Method \\
			\midrule
			\cite{leeSplittableDNNBasedObject2021a} & Transformed & Feature reconstructor \\
			ADCNN \cite{zhangAdaptiveDistributedConvolutional2020a} & Transformed & FDSP \\
			EDDL \cite{changEfficientDistributedDeep2019a} & Transformed & Grouped convolutions\\
			Capella \cite{baeCapellaCustomizingPerception2019a} & Transformed & Eliminate synchronization points \\
			CNNPC \cite{yangCNNPCEndEdgeCloudCollaborative2022} & Transformed & ICP + CRD \\
			\cite{zhangCommunicationComputationEfficientDeviceEdge2021} & Transformed & AE \\
			\cite{shaoCommunicationComputationTradeoffResourceConstrained2020} & Transformed & AE \\
			\cite{yunCooperativeInferenceDNNs2022a} & New & KD \\
			CDE \cite{sbaiCutDistilEncode2021a} & New & KD + AE \\
			DPDS \cite{zhouDynamicPathBased2021a} & Transformed & Early exit branches \\
			\cite{huFastAccurateStreaming2020a} & Transformed & AE \\
			EdgeDI \cite{fangJointArchitectureDesign2022a} & Transformed & SCAR \\
			\cite{jiangJointModelTask2022a} & Transformed & Early exit branches \\
			MAHPPO \cite{haoMultiAgentCollaborativeInference2022} & Transformed & AE \\
			\cite{jeongOptimalPartitioningDistributed2021a} & New & KD \\
			TeamNet \cite{fangTeamNetCollaborativeInference2019a} & New & Expert models\\
			\cite{teerapittayanonDistributedDeepNeural2017} & Transformed & Early exit branches\\
			\cite{matsubaraDistilledSplitDeep2019} & New & KD \\
			BottleNet++ \cite{shaoBottleNetEndtoEndApproach2020} & Transformed & AE \\
			BottleNet \cite{eshratifarBottleNetDeepLearning2019} & Transformed & AE \\
			\cite{luoCloudEdgeCollaborativeIntelligent2021} & Transformed & Early exit branches\\
			\bottomrule
		\end{tabularx}
		\label{tab:resulting_nn_topology}
	\end{center}
\end{table}
}

\subsubsection{Embedded devices used}
\label{section:devices_used}


One important distinction between papers pertains to the embedded device they target. It is important to analyse and keep in mind the computing capabilities of the modelled devices, as they can directly influence the decisions taken by the distribution algorithm. As an example, we can mention the case of optimizing for latency or energy for a system composed of devices whose computing capabilities are far superior to the capacity of the communication channels between them. In this case, an algorithm which is aware of the network's parameters should be used, because the cost of exchanging data between devices overshadows the cost of the execution of the network. If an algorithm that does not integrate information about the communication channel is used, the resulting distributed solution can be sub-optimal, as the gain of optimizing the execution of the network is negligible when compared with the cost of moving data across devices.

It is also important to categorize distributed inference papers across what kind of device is targeted, as it gives a good overview of the current focus of the area, and helps identify gaps in the SotA implementations. Table \ref{tab:devices_used} provides our categorization regarding embedded devices targeted by the papers that were surveyed. As can be seen, most papers use Raspberry Pi devices or similar development boards. Although these platforms are useful to test new algorithms and concepts, other papers use HW that is more AI-specific, like embedded GPUs (usually from the NVIDIA Jetson family). A number of papers have focused on mobile devices (smartphones). Others focused on generic CPUs cores (including distributing across multiple MCUs). 

{\footnotesize
\begin{table}[htbp]
	\caption{Papers categorized according to the kind of embedded devices they used in their experiments/simulations}
	\begin{center}
		\begin{tabularx}{\textwidth}{>{\hsize=.15\hsize}X>{\hsize=.8\hsize}X} \toprule
			Category & Papers \\
			\midrule
			Raspberry Pi kind & \cite{wangFastEnergySavingNeural2021,zhangAdaptiveDistributedConvolutional2020a,miaoAdaptiveDNNPartition2020a,samikwaAdaptiveEarlyExit2022a,zhouAdaptiveParallelExecution2019a,niuAdaptiveDeviceEdgeCoInference2022,zengBoomerangOnDemandCooperative2019,zengCoEdgeCooperativeDNN2021a,zhangCommunicationComputationEfficientDeviceEdge2021,shaoCommunicationComputationTradeoffResourceConstrained2020,yangCooperativeDistributedDeep2021a,yunCooperativeInferenceDNNs2022a,zhangDeepSlicingCollaborativeAdaptive2021a,zhaoDeepThingsDistributedAdaptive2018a,houDistrEdgeSpeedingConvolutional2022a,gacoinDistributingDeepNeural2019a,huDynamicAdaptiveDNN2019a,zhangDynamicDNNDecomposition2021a,DynamicSplitComputing2022,liEdgeAIOnDemand2020,huEnablePipelineProcessing2021a,huangEnablingLowLatency2021a,huFastAccurateStreaming2020a,wuHiTDLHighThroughputDeep2022,fangJointArchitectureDesign2022a,fuJointOptimizationData2021a,dongJointOptimizationDNN2021a,liuLoADPartLoadAwareDynamic2022,naveenLowLatencyDeep2021a,deyOffloadedExecutionDeep2019a,deyPartitioningCNNModels2018a,shiPrivacyAwareEdgeComputing2019,baccourRLPDNNReinforcementLearning2021a,SplitComputingDNN2022,tuliSplitPlaceAIAugmented2022a,fangTeamNetCollaborativeInference2019a,yangEfficientInferenceAdaptively2021a,zhangRealTimeCooperativeDeep2020,baccourDistPrivacyPrivacyAwareDistributed2020,hadidiDistributedPerceptionCollaborative2018a,hadidiCollaborativeInferencingDeep2020a,zhouAAIoTAcceleratingArtificial2019,wangADDAAdaptiveDistributed2019,matsubaraDistilledSplitDeep2019} \\
			Embedded GPUs & \cite{leeSplittableDNNBasedObject2021a,zhangAutodidacticNeurosurgeonCollaborative2021a,dagliAxoNNEnergyAwareExecution2022,yangCNNPCEndEdgeCloudCollaborative2022,zengCoEdgeCooperativeDNN2021a,vanishreeCoInAcceleratedCNN2020a,houDistrEdgeSpeedingConvolutional2022a,liDistributedDeepLearning2022a,almeidaDynODynamicOnloading2022a,zhangElfAccelerateHighResolution2021,huangEnablingLowLatency2021a,xuEOPEfficientOperator2022a,kressHardwareawarePartitioningConvolutional2022a,wuHiTDLHighThroughputDeep2022,liJALADJointAccuracyAnd2018,haoMultiAgentCollaborativeInference2022,yangOffloadingOptimizationEdge2021a,liReceptiveFieldbasedSegmentation2022a,fangTeamNetCollaborativeInference2019a,eshratifarJointDNNEfficientTraining2021,kangNeurosurgeonCollaborativeIntelligence2017a,laskaridisSPINNSynergisticProgressive2020,matsubaraDistilledSplitDeep2019,eshratifarBottleNetDeepLearning2019} \\
			Mobile & \cite{kimAutoScaleEnergyEfficiency2020a,heidariCAMDNNContentAwareMapping2022,yangCNNPCEndEdgeCloudCollaborative2022,jiaCoDLEfficientCPUGPU2022a,huangDeeParHybridDeviceEdgeCloud2019,DynamicSplitComputing2022,liuEEAIEndedgeArchitecture2021a,zhangElfAccelerateHighResolution2021,liEnablingRealtimeAI2022a,parasharProcessorPipeliningMethod2020a,SplitComputingDNN2022,baccourDistPrivacyPrivacyAwareDistributed2020,xuDeepWearAdaptiveLocal2020,maoMoDNNLocalDistributed2017a,renEdgeAssistedDistributedDNN2020,changUltraLowLatencyDistributedDeep2019} \\
			CPUs & \cite{changEfficientDistributedDeep2019a,vanishreeCoInAcceleratedCNN2020a,sbaiCutDistilEncode2021a,xueDDPQNEfficientDNN2022a,sahuDENNIDistributedNeural2021a,zhouDynamicPathBased2021a,wangDynamicResourceAllocation2021a,xueEdgeLDLocallyDistributed2020a,jiangJointModelTask2022a,choiLegionTailoringGrouped2021a,jiNovelAdaptiveDNN2022a,luResourceEfficientDistributedDeep2022,camposdeoliveiraPartitioningConvolutionalNeural2018} \\
			Accelerator & \cite{banitalebi-dehkordiAutoSplitGeneralFramework2021,zouCAPCommunicationAwareAutomated2022,jinEnergyAwareWorkloadAllocation2019,koEdgeHostPartitioningDeep2018,huangCLIOEnablingAutomatic2020} \\
			FPGAs & \cite{jiangAchievingSuperLinearSpeedup2019,baeCapellaCustomizingPerception2019a,vanishreeCoInAcceleratedCNN2020a} \\
			Wearable & \cite{baccourDistPrivacyPrivacyAwareDistributed2020,xuDeepWearAdaptiveLocal2020} \\
			\bottomrule
		\end{tabularx}
		\label{tab:devices_used}
	\end{center}
\end{table}
}

\subsubsection{Open sourced projects}
\label{section:open_source}

The reproducibility of any computer science paper can be greatly improved when the code used for the generation of the experimental results is opened to the research community. As such, it is concerning that only 14 \% of the surveyed papers have published their code (see Table \ref{tab:open_source} \footnote{Two papers \cite{kimAutoScaleEnergyEfficiency2020a,huangCLIOEnablingAutomatic2020} provide links that do not work any more.}). Others \cite{xuDeepWearAdaptiveLocal2020} say they are interested in publishing their code, but the authors of this paper were not able to find any open implementation for them. 

{\footnotesize
\begin{table}[htbp]
	\caption{Links to the open source releases of the reviewed projects}
	\label{tab:open_source}
	\begin{center}
		\begin{tabular}{cc} \toprule
			Paper & URL \\
			\midrule
			Auto-Split \cite{banitalebi-dehkordiAutoSplitGeneralFramework2021} & \url{https://github.com/abanitalebi/auto-split}\\ 
			Capella \cite{baeCapellaCustomizingPerception2019a} & \url{https://github.com/parallel-ml/Capella-FPL19-SplitNetworksOnFPGA}\\ 
			CNNPC \cite{yangCNNPCEndEdgeCloudCollaborative2022} & \url{https://github.com/IoTDATALab/CNNPC}\\ 
			\cite{shaoCommunicationComputationTradeoffResourceConstrained2020} & \url{https://github.com/shaojiawei07/Edge_Inference_three-step_framework}\\ 
			DeepThings \cite{zhaoDeepThingsDistributedAdaptive2018a} & \url{https://github.com/SLAM-Lab/DeepThings}\\ 
			DEFER \cite{parthasarathyDEFERDistributedEdge2022a} & \url{https://github.com/anrgusc/defer}\\ 
			EdgeLD \cite{xueEdgeLDLocallyDistributed2020a} & \url{https://github.com/fangvv/EdgeLD}\\ 
			Elf \cite{zhangElfAccelerateHighResolution2021} & \url{https://github.com/wuyangzhang/elf}\\ 
			POPEX \cite{pachecoInferenceTimeOptimization2020a} & \url{https://github.com/pachecobeto95/POPEX}\\ 
			EdgeDI \cite{fangJointArchitectureDesign2022a} & \url{https://github.com/fangvv/EdgeDI}\\ 
			MAHPPO \cite{haoMultiAgentCollaborativeInference2022} & \url{https://github.com/Hao840/MAHPPO}\\ 
			DPFP \cite{liReceptiveFieldbasedSegmentation2022a} & \url{https://gitlab.au.dk/netx/agileiot/dpfp}\\ 
			SplitPlace \cite{tuliSplitPlaceAIAugmented2022a} & \url{https://github.com/imperial-qore/SplitPlace}\\ 
			\cite{teerapittayanonDistributedDeepNeural2017} & \url{https://github.com/kunglab/ddnn}\\ 
			\cite{matsubaraDistilledSplitDeep2019} & \url{https://github.com/yoshitomo-matsubara/head-network-distillation}\\ 
			BootleNet++ \cite{shaoBottleNetEndtoEndApproach2020} & \url{https://github.com/shaojiawei07/BottleNetPlusPlus}\\ 
			\bottomrule
		\end{tabular}
	\end{center}
\end{table}
}

\subsubsection{Trends and challenges}

Although Table \ref{tab:dataset_used} shows that datasets like ImageNet dominate the distributed inference landscape, datasets like MNIST \cite{lecun-mnisthandwrittendigit-2010}, CIFAR-10 \cite{krizhevsky2009learning} have also been widely used. But their small image dimensions do not provide a particularly interesting challenge for the distribution of the DNN across multiple devices. This happens because the intermediate feature maps of the DNN have also small dimensions, which reduces the importance and impact of the bandwidth between devices. Bigger feature maps provide a more challenging setting for the partitioning algorithm, by forcing it to try to fuse more layer executions to reduce the amount of data exchanged between devices. This is why we would like to encourage the reader to consider using datasets that provide bigger images (for example, ImageNet) in future distributed inference papers. 

In terms of the used DNN architectures, VGG is still the most commonly distributed DNN, but given the rising popularity of vision transformers models, we expect transformers to dominate this list in the following years.

By analysing Table \ref{tab:resulting_nn_topology}, although KD is a useful approach to reduce the complexity of the original network, which should help the distribution of the new network, it must be taken into account that this requires training a completely new DNN. Something similar happens with the use of AE: both the encoder and decoder networks need to be trained to avoid hurting the performance of the prediction of the original network. For AE, it is also important to analyse the trade-off between the reduction of the size of the transmitted data and the additional cost inserted because of the execution of the encoder/decoder networks. 

In terms of the devices that are targets of the distribution algorithms, a promising field for future studies is the distribution across CPU cores and multi-accelerator architectures, which have not yet received as much attention as typical consumer devices like RPi. Multi-FPGA systems and systems composed of wearable-mobile devices are also promising fields on which distributed inference papers can be deployed.

Finally, we would like to encourage the reader to keep in mind the benefits of open sourcing implementations when publishing a paper on this topic, as this could greatly improve the quality of future research.



\section{Conclusions}
\label{section:conclusions}

This comprehensive survey provided a thorough examination of SotA papers on embedded distributed inference. We effectively categorized and discussed the design approaches used in these systems by examining current trends and introducing a novel taxonomy. Furthermore, as outlined in the different "Trends and challenges" sections throughout this work, we have highlighted the existing challenges and limitations that surround the field. As the popularity of this topic grows, it is clear that there are several unanswered questions that need to be addressed. By providing this review, we hope to present a valuable resource for future researchers working on distributed inference systems for AI.

	
	\bibliographystyle{ACM-Reference-Format}
	\bibliography{bib}
	
	
\end{document}